%% file: draft_VtoN.tex
\documentclass[twocolumn,prd,noshowpacs,nofootinbib,amsmath,amssymb,superscriptaddress,preprintnumbers]{revtex4}
\pdfoutput=1
\usepackage{graphicx,epsfig,psfrag,bm,amssymb}
\usepackage{dcolumn}
\usepackage{bm}
\usepackage{mciteplus}
\usepackage{color}
\usepackage{mathrsfs,amsfonts,hepunits, color}
\usepackage{mciteplus}
\usepackage{tikz}
\usepackage{slashed}
\usepackage{cancel}
\usetikzlibrary{arrows,shapes}
\usetikzlibrary{trees}
\usetikzlibrary{matrix,arrows} 				
\usetikzlibrary{positioning}				
\usetikzlibrary{calc,through}				
\usetikzlibrary{decorations.pathreplacing}  
\usepackage{pgffor}							

\usetikzlibrary{decorations.pathmorphing}	
\usetikzlibrary{decorations.markings}
\usetikzlibrary{snakes}

\tikzset{
    vector/.style={decorate, decoration={snake}, draw},
	provector/.style={decorate, decoration={snake,amplitude=2.5pt}, draw},
	antivector/.style={decorate, decoration={snake,amplitude=-2.5pt}, draw},
    fermion/.style={draw, postaction={decorate},
        decoration={markings,mark=at position .55 with {\arrow[draw]{>}}}},
    fermionbar/.style={draw, postaction={decorate},
        decoration={markings,mark=at position .55 with {\arrow[draw=black]{<}}}},
    fermionnoarrow/.style={draw},
    gluon/.style={decorate, draw,decoration={coil,amplitude=4pt, segment length=6pt}, line width=1},
    scalar/.style={dashed,draw, postaction={decorate},
        decoration={markings,mark=at position .55 with {\arrow[draw]{>}}}},
    scalarbar/.style={dashed,draw, postaction={decorate},
        decoration={markings,mark=at position .55 with {\arrow[draw]{<}}}},
    scalarnoarrow/.style={dash pattern = on 6 pt off 3 pt,draw},
    electron/.style={draw, postaction={decorate},
        decoration={markings,mark=at position .55 with {\arrow[draw]{>}}}},
	bigvector/.style={decorate, decoration={snake,amplitude=4pt}, draw},
	vectorscalar/.style={loosely dotted,draw, postaction={decorate}},
}
\RequirePackage{xspace}
\usepackage{relsize}
\usepackage{slashed}

\newcommand{\be}{\begin{eqnarray}}
\newcommand{\ee}{\end{eqnarray}}
\def\lsim{\mathrel{\rlap{\lower4pt\hbox{\hskip 0.5 pt$\sim$}}
    \raise1pt\hbox{$<$}}}                
\def\gsim{\mathrel{\rlap{\lower4pt\hbox{\hskip1pt$\sim$}}
    \raise1pt\hbox{$>$}}} 
    
\def\lsim{\mathrel{\rlap{\lower4pt\hbox{\hskip1pt$\sim$}}
    \raise1pt\hbox{$<$}}}
\def\gsim{\mathrel{\rlap{\lower4pt\hbox{\hskip1pt$\sim$}}
    \raise1pt\hbox{$>$}}}

\newcommand{\bi}{\begin{itemize}}
\newcommand{\ei}{\end{itemize}}
\newcommand{\bea}{\begin{eqnarray}}
\newcommand{\eea}{\end{eqnarray}}

\newcommand{\benum}{\begin{enumerate}}
\newcommand{\eenum}{\end{enumerate}}

\begin{document}

\preprint{PITT-PACC-1604}
\preprint{SLAC-PUB-16511}

\title{ 
Shedding Light on Neutrino Masses with Dark Forces}
\author{Brian Batell}
\affiliation{Pittsburgh Particle Physics, Astrophysics, and Cosmology Center, Department of Physics and Astronomy, University of Pittsburgh, PA 15260, USA}
\author{Maxim Pospelov}
\affiliation{Perimeter Institute for Theoretical Physics, Waterloo, ON N2J 2W9, Canada}
\affiliation{Department of Physics and Astronomy, University of Victoria, Victoria, BC V8P 5C2, Canada}
\author{Brian Shuve}
 \affiliation{SLAC National Accelerator Laboratory, 2575 Sand Hill Road, Menlo Park, CA 94025, USA}

\begin{abstract}

Heavy right-handed neutrinos, $N$, provide the 
simplest explanation for the origin of light neutrino masses and mixings. 
If $M_N$ is at or below the weak scale, direct experimental discovery of these states is possible at accelerator experiments such as the LHC or new dedicated beam dump experiments; in these experiments,
$N$ decays after traversing a macroscopic distance from the collision point. 
The experimental sensitivity to right-handed neutrinos is significantly enhanced if there is a new ``dark'' gauge force connecting them to the Standard Model (SM), and detection of $N$ can be the primary discovery mode for the new dark force itself. 
We take the well-motivated example of a $B-L$ 
gauge symmetry and analyze the sensitivity to displaced decays of $N$ produced via the new gauge interaction in two experiments:~the LHC and the proposed SHiP beam dump experiment. 
In the most favorable case in which the mediator can be produced on-shell and decays to right handed neutrinos ($pp\to X + V_{B-L}\to X+ N N$), 
the sensitivity reach is controlled by the square of the $B-L$ gauge coupling.
We demonstrate that these experiments could access neutrino parameters responsible for the observed SM neutrino masses and mixings in the most straightforward 
implementation of the 
see-saw mechanism. 
\end{abstract}

\maketitle

%
%
\section{Introduction}
\label{sec:intro}

\input{introduction}

\section{Right-Handed Neutrinos and New Gauge Forces}\label{sec:simplified_model}


\input{simplified_model}

\section{LHC Sensitivity to $N$ From Vector Decay}\label{sec:LHC}

\input{LHC}

\section{SHiP Sensitivity to RH Neutrinos}\label{sec:SHiP}

\input{SHiP}

\section{Discussion and Conclusions}\label{sec:concl}

\input{concl}

\bibliography{biblio_VtoN}

\end{document}

%% file: introduction.tex
Since the first discovery of neutrino oscillations over fifteen years ago~\cite{Fukuda:1998mi,Ashie:2004mr,Fukuda:2002pe,Ahmad:2002jz,Eguchi:2002dm}, neutrino masses and mixings have been hailed as the first definitive evidence from particle physics experiments of physics beyond the Standard Model (SM). Understanding the physics of SM neutrino masses may therefore shed light on other unsolved problems in fundamental physics, such as dark matter or the baryon asymmetry.  From the perspective of effective field theory, neutrino masses can be incorporated in the SM via the dimension-5 Weinberg operator, $ c(LH)^2 / \Lambda$ \cite{Weinberg:1979sa}, where the cutoff $\Lambda$ could range anywhere from $10^{-9}-10^{16}$ GeV depending on the coupling $c$. It is evident that the new fields responsible for neutrino masses could appear at a wide range of scales, and it is imperative that models of neutrino mass generation are tested in as broad a manner as possible by available experiments.

In the SM, all left-handed (LH) charged fermions acquire a Dirac mass by coupling to the Higgs and a corresponding right-handed (RH) field. If the LH neutrinos acquire Dirac masses $M_{\rm D}$ through the same mechanism, the SM must be supplemented with RH neutrinos (RHNs), $N$, which in the simplest case of a type-I seesaw are singlets with respect to the SM gauge interactions. As singlets, the $N$ fields can have arbitrary Majorana masses, $M_N$; in the limit $M_N\gg M_{\rm D}$, this scenario provides the most natural ultraviolet (UV) completion of the Weinberg operator above. After electroweak symmetry breaking, the neutrino mass matrix is not diagonal; in the simplified case of one LH and one RH neutrino, the mass eigenstates are
\bea
m_\nu &\approx& \frac{M_{\rm D}^2}{M_N} ,\label{eq:seesaw}\\
M &\approx& M_N,
\eea
where $m_\nu$ is the observed SM neutrino mass and $M$ is the mass of a new heavy state.  The SM neutrino masses are suppressed by the heavy Majorana scale, and this is the most straightforward implementation of the see-saw mechanism \cite{Minkowski:1977sc,Yanagida:1979as,Mohapatra:1979ia,GellMann:1980vs,Schechter:1980gr}\footnote{In the see-saw limit, $M$ and $M_N$ can be used interchangeably, and from now on we use only $M_N$.}.

The neutrino mass eigenstates are not completely aligned with the lepton doublet and singlet fields; the light SM-like neutrino mass eigenstate acquires a small component of the singlet, and the heavy singlet-like state acquires a small coupling under the weak interactions. The mixing angle, $\theta$, between the neutrino states is (in the see-saw limit)
\be\label{eq:theta_def}
\theta \approx \frac{M_{\rm D}}{M_N},
\ee
and $\theta$ determines how strongly the sterile RH neutrino $N$ couples to the SM. Indeed, the matrix element for any process coupling $N$ to SM fields is the same as the corresponding coupling of LH neutrinos to the SM, multiplied by a factor of $\theta$. Using Eq.~(\ref{eq:seesaw}), one finds
\be\label{eq:mixing}
\theta^2 \approx \frac{m_\nu}{M_N};
\ee 
the larger the $N$ mass, the more weakly coupled it is to the SM to explain the observed LH neutrino masses.

The scale of $m_\nu$ is not measured directly, as neutrino oscillation experiments probe only the squared mass splittings, $\Delta m_\nu^2$. The actual values of $m_\nu$ can vary from massless (which is a viable option only for the lightest mass eigenstate) to the upper bounds supplied by cosmology ($m_\nu\lesssim 0.23$ eV)  \cite{Ade:2015xua} and direct neutrino mass searches, ($m_{\nu_e}\lesssim 2\,\,\mathrm{eV}$) \cite{Aseev:2011dq}. For the heavier mass eigenstates, a lower bound is given by the experimentally determined squared mass splittings. For both the normal and inverted hierarchy at least one mass eigenstate must be heavier than $\sqrt{\Delta (m_\nu^2)^{\rm atm}} \simeq 0.05\,\,\mathrm{eV}$, giving a lower bound on the mixing angle. 
From the see-saw relation in Eq.~(\ref{eq:mixing}), 
the expected value of the mixing angle is: 
\be
\label{thetass}
\theta_{\rm s-s}^2 \sim 5\times 10^{-11} \times \left( \frac{1~{\rm GeV}}{M_N} \right).
\ee
This represents a well-motivated target for experimental searches for right-handed neutrinos. It must be emphasized, however,  that more complicated mass generation schemes could produce significantly larger or smaller  $\theta_{\rm s-s}$~\cite{Alekhin:2015byh}\footnote{In particular, $M_D$ and therefore $\theta$ are in fact complex matrices, and a cancellation between real and imaginary parts can result in $\theta^{\rm T}\theta \ll \theta^\dagger \theta$; in other words, the mixing angles can be much larger than na\"ively expected by Eq.~(\ref{thetass}). This occurs in models with approximate lepton number conservation \cite{Casas:2001sr,Shaposhnikov:2006nn} such as the inverse see-saw \cite{Mohapatra:1986bd}.}.

The mass of the heavy, sterile state $M_N$ is essentially a free parameter of the model. 
Of particular interest to us are masses that are kinematically accessible to current experiments, $M_N\lesssim$ TeV; the RH neutrino can be directly produced in SM interactions, but the production rate scales like $|\theta|^2$. In this mass range, Eq.~(\ref{thetass}) suggests that the RH neutrinos are  produced in SM interactions only very rarely,  making the see-saw mechanism very difficult to test in direct experiments.
Current sensitivity to $\theta_{\rm s-s}$ only exists in the window of $1\,\,\mathrm{MeV}$ to a few hundred MeV, in which $\theta_{\rm s-s}$ 
is strongly disfavored by the combination of Big Bang Nucleosynthesis (BBN) and cosmic microwave background (CMB) data \cite{Ruchayskiy:2012si}.

\begin{figure}[t]
\centering
\includegraphics[width=0.3 \textwidth ]{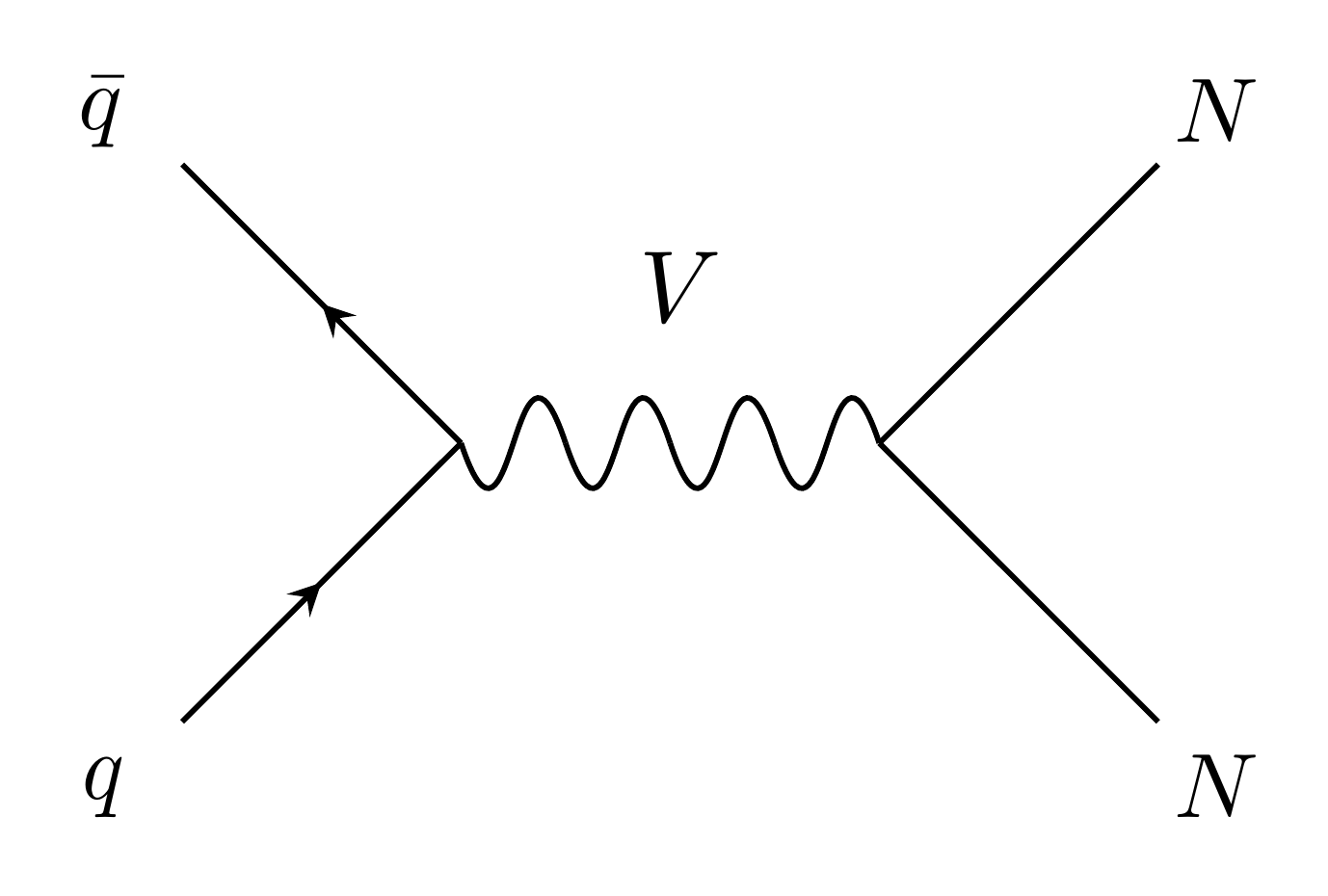}
\caption{Production of right-handed neutrinos, $N$, via a new gauge interaction at hadron colliders or proton beam dumps.
}
\label{fig:N_pair_prod}
\end{figure}

The prospects for discovering RHNs satisfying Eq.~(\ref{thetass}) are significantly improved if they can be produced through  interactions other than the mixing angle $\theta$. For example, if the RHN and SM fields are both charged under a new ``dark force'', then $N$ pairs can be produced via this gauge interaction independently of the value of $\theta$ \cite{delAguila:2007ua,Huitu:2008gf,Basso:2008iv,Blanchet:2009bu,Perez:2009mu,Kang:2015uoc,Okada:2016gsh},  as shown in Fig.~\ref{fig:N_pair_prod}\footnote{In other models, RHN can also be pair produced via a new scalar \cite{Shoemaker:2010fg} or singly produced via a new right-handed $W$ boson \cite{Keung:1983uu}.}. Indeed, this coupling of $N$ to the dark force is mandatory in the simplest gauge extension of the SM, in which the SM is supplemented by a new $\mathrm{U}(1)_{B-L}$ local symmetry \cite{Mohapatra:1980qe} with coupling $g'$ and vector boson $V$; anomaly cancelation requires the extension of the SM with three additional RHNs. Because ${g'}^2$ can exceed $|\theta|^2$ by many orders of magnitude, the new gauge interaction allows for the discovery of $N$ even for the tiny mixing angles predicted by Eq.~(\ref{thetass}).

\begin{figure}[t]
\centering
\includegraphics[width=0.5 \textwidth ]{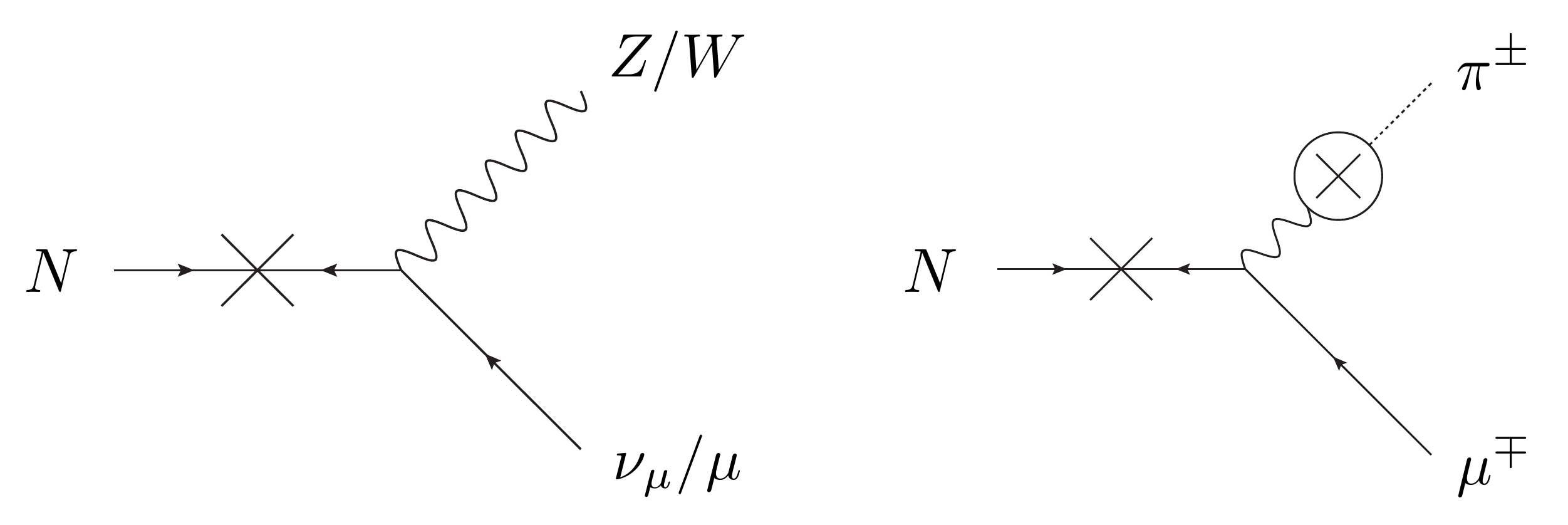}
\caption{\emph{(Left):}~Right-handed neutrinos ($N$) decay via the electroweak interactions  due to mixing with LH neutrinos; they also decay to the Higgs via Yukawa couplings (not shown). \emph{(Right)}:~At low masses, $M_N\lesssim$ GeV, the exclusive hadronic decays of $N$, such as $N\rightarrow \pi^\pm \mu^\mp$, are relevant.}
\label{fig:N_decay}
\end{figure}

Although $N$ can be pair produced through new gauge interactions at colliders and beam-dump experiments, the RHNs can only decay through its tiny mixing with SM neutrinos (see Fig.~\ref{fig:N_decay}); consequently, the $N$ width is expected to be very small. For RHN masses within range of current colliders,  $M_N \lesssim 200$ GeV,
the decays of $N$ occur on macroscopic distance scales for mixing angles consistent with Eq.~(\ref{thetass}) \cite{Basso:2008iv,Perez:2009mu}. This gives rise to spectacular 
signatures at accelerator experiments, such as displaced vertices at the Large Hadron Collider (LHC) and visible decays of $N$ at the new planned SHiP facility  \cite{Anelli:2015pba,Alekhin:2015byh}. We perform here a quantitative study of the possible long-lived particle searches that have sensitivity to RHNs with a new dark force\footnote{Displaced vertex searches have also been found to be useful in discovering RHNs produced via mixing with LH neutrinos at the LHC \cite{Helo:2013esa,Izaguirre:2015pga} and future colliders \cite{Blondel:2014bra,Antusch:2016vyf}.}. In addition to enhancing the detection prospects for RHN that would otherwise be out of reach of direct experimental probes, the sensitivity of the LHC and SHiP to long-lived particle signatures is sufficiently good that the process $pp\rightarrow V\rightarrow NN$ can serve as the primary discovery mode of the new $\mathrm{U}(1)$ gauge interaction. For concreteness, we focus on the well-motivated case of a $B-L$ gauge symmetry, but many of our conclusions can be carried over to other examples.

\begin{figure}[t]
\centering
\includegraphics[width=0.48 \textwidth ]{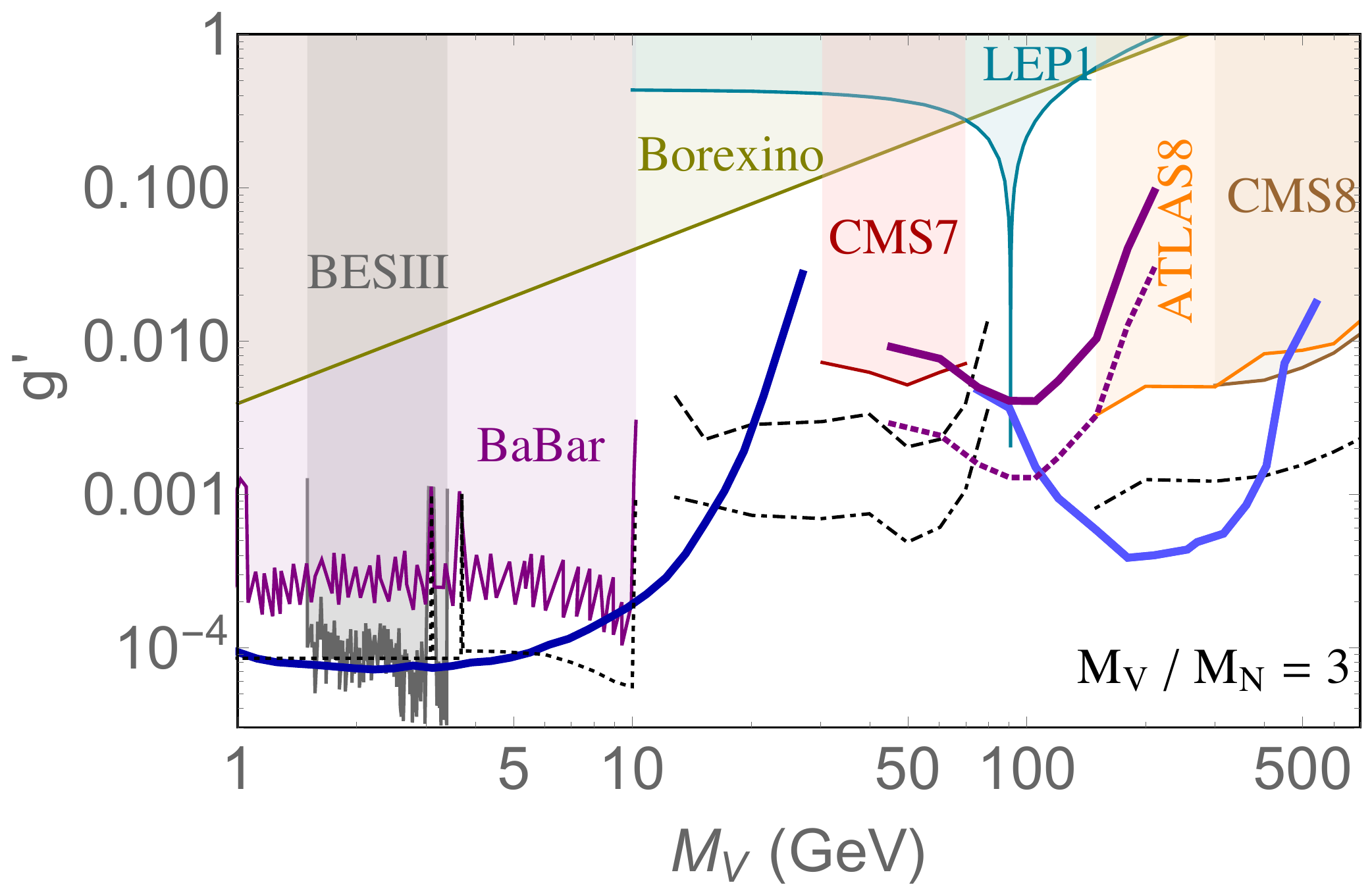}
\caption{Current constraints and future sensitivity to the $\mathrm{U}(1)_{B-L}$ model with $M_V/M_N=3$. The shaded regions are excluded by the indicated experiment. The projected reach of our proposed searches for $V_{B-L}\rightarrow NN$ are shown in thick curves from SHiP (left, dark blue) and the high-luminosity LHC ($3\,\,\mathrm{ab}^{-1}$):~inner-detector displaced vertex search (light blue) and muon spectrometer displaced vertex search (purple; solid for high background scenario, dashed for low background).  The RH neutrino mixing angle is fixed using Eq.~(\ref{thetass}). The thin black curves show the projected sensitivity of direct searches for $V_{B-L}\rightarrow\ell^+\ell^-$ from Belle II (dotted), LHC Run 1 (dashed), and the high-luminosity LHC (dot-dashed).
}
\label{fig:master_MV_gp}
\end{figure}

\begin{figure}[t]
\centering
\includegraphics[width=0.48 \textwidth ]{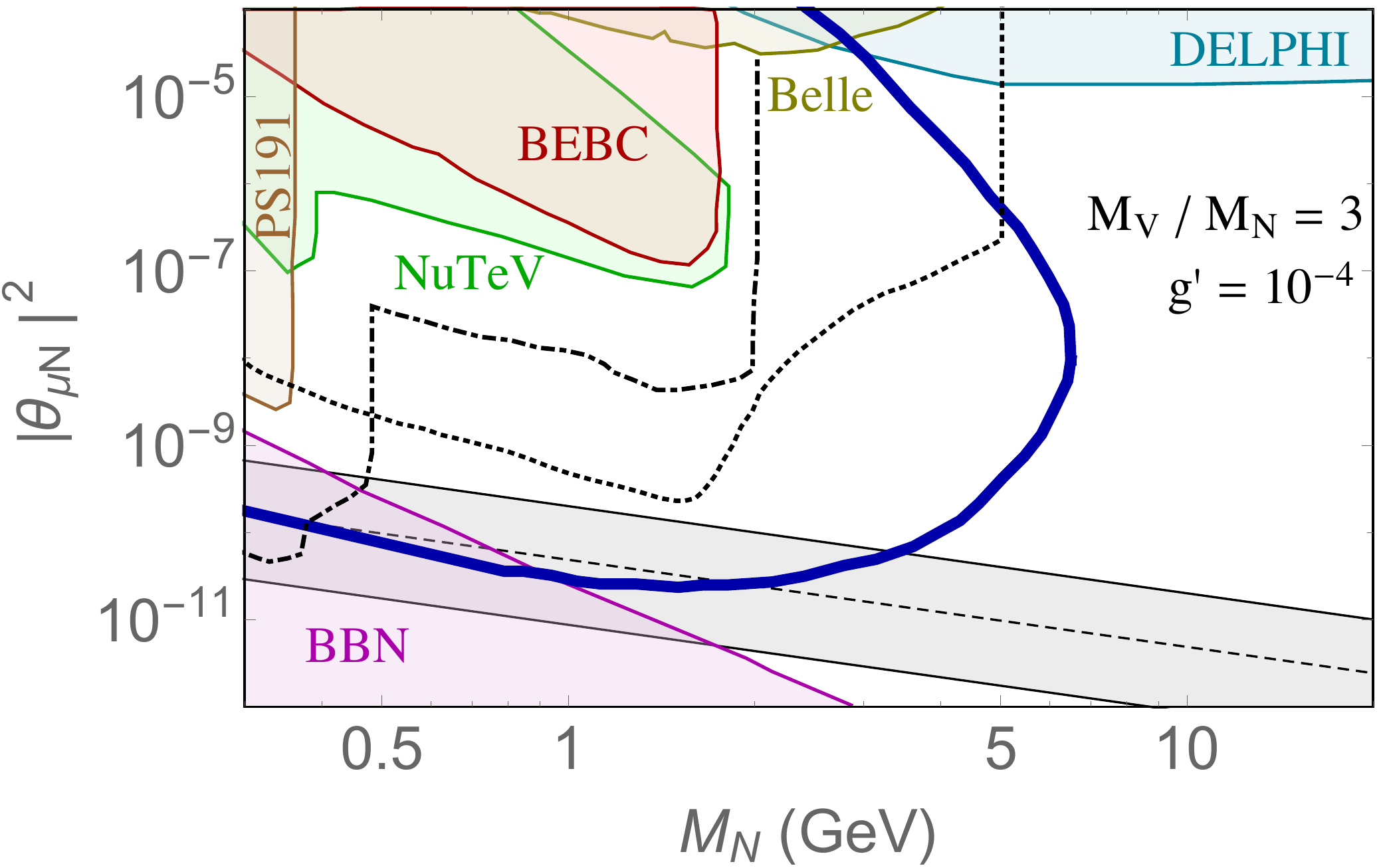}
\caption{Current constraints and future sensitivity to right-handed neutrinos in the $\mathrm{U}(1)_{B-L}$ model with $M_V/M_N=3$ and $g'=10^{-4}$. The shaded regions are excluded by the indicated experiment. The thick blue curve shows the projected reach of a SHiP search for $N$ production in $V_{B-L}\rightarrow NN$, while the thin dashed line shows the SHiP sensitivity to direct $N$ production through its mixing with LH neutrinos.  The thin dot-dashed curve shows the sensitivity for a near detector at DUNE to direct $N$ production \cite{Adams:2013qkq}. The shaded grey band is the region preferred by the see-saw mechanism; see Fig.~\ref{fig:mixing_lifetime} for more details. 
}
\label{fig:master_MN_V2_SHiP}
\end{figure}

\begin{figure}[t]
\centering
\includegraphics[width=0.48 \textwidth ]{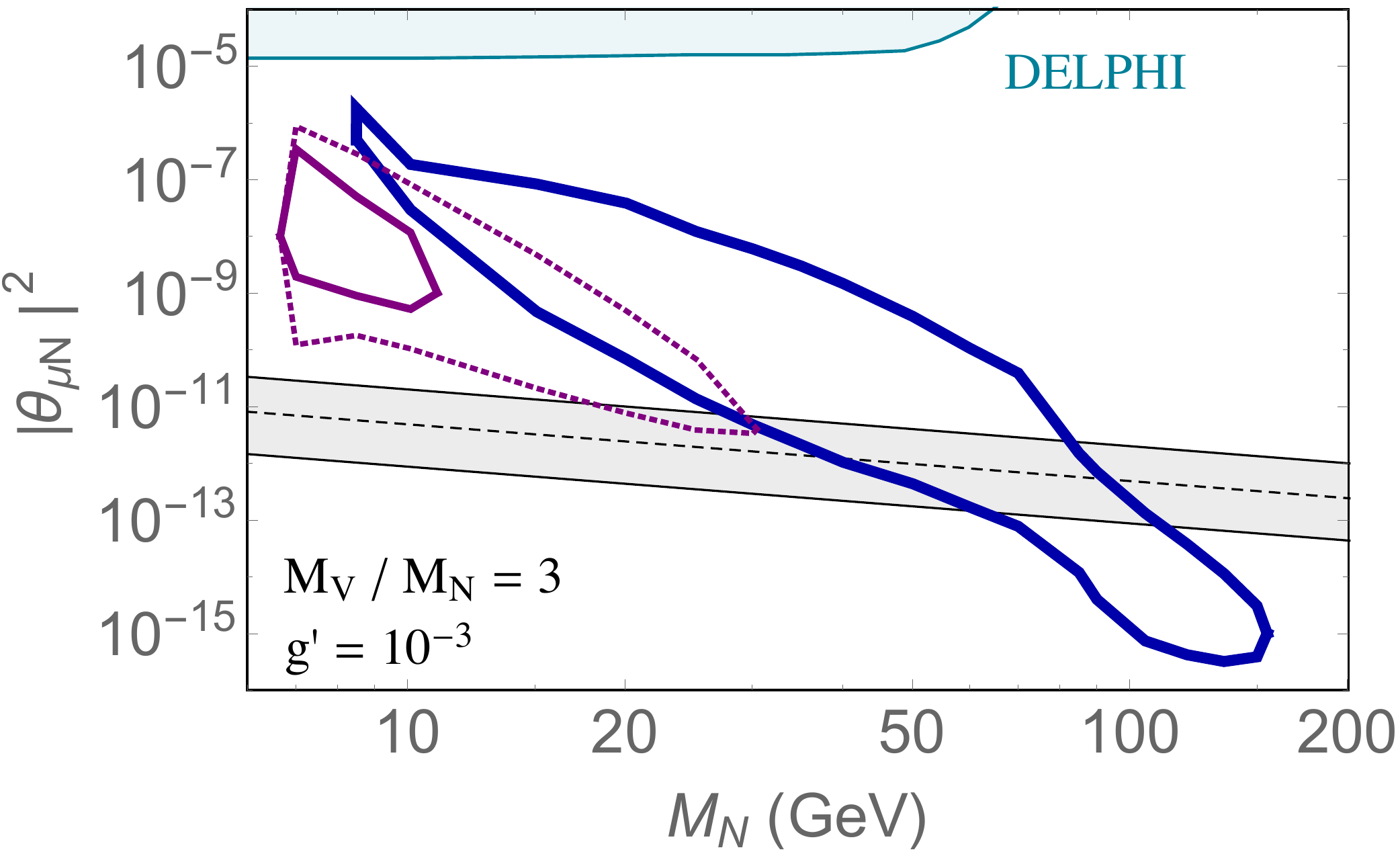}
\caption{Current constraints and future sensitivity to right-handed neutrinos in the $\mathrm{U}(1)_{B-L}$ model with $M_V/M_N=3$ and $g'=10^{-3}$. The shaded regions are excluded by the indicated experiment. The thick light blue curve shows the projected reach at the high-luminosity LHC ($3\,\,\mathrm{ab}^{-1}$) of our proposed searches for displaced vertices in the inner detector from $V_{B-L}\rightarrow NN$, while the purple curves show sensitivity for a search for displaced vertices in the muon spectrometer (solid for high background scenario, dashed for low background).  The shaded grey band is the region preferred by the see-saw mechanism; see Fig.~\ref{fig:mixing_lifetime} for more details. 
}
\label{fig:master_MN_V2_LHC}
\end{figure}

Jumping ahead to the results of our study, we show current constraints and projected future sensitivity from the high-luminosity LHC and SHiP to the $B-L$ model 
with RHNs in Figures \ref{fig:master_MV_gp}, \ref{fig:master_MN_V2_SHiP} and \ref{fig:master_MN_V2_LHC}. These figures show 
that sensitivity to both a new $B-L$ force and RHN mixing parameters are poised to significantly improve in coming years. In particular, 
both the high-luminosity LHC and SHiP searches will be able to directly explore parts of the parameter space motivated by the see-saw mechanism.

This paper is organized as follows:~in the next section we introduce scenarios with a new gauge force and discuss its broad impact on the phenomenology of 
$N$. In section 3, we consider the pair production of $N$ at the LHC and estimate the sensitivity to the doubly-displaced decays of $N$, 
comparing our results to the constraints on $V$ that can be derived from its direct decays into SM particles. 
In section 4, we deduce the sensitivity to $N$ at SHiP via the production of $V$ in proton collisions at a beam dump, followed by the visible decays of $N$ in a detector far downstream from the beam dump. 
We reach our conclusions in section 5. 

%% file: simplified_model.tex
The SM admits several possibilities for an additional $\mathrm{U}(1)'$ gauge force and its associated gauge boson, $V$; this is often called the ``vector portal'' or a ``dark force''. 
The most discussed SM extension in this category is the ``kinetic mixing'' coupling, 
$\epsilon V_{\mu\nu} B^{\mu\nu}/2$ \cite{Holdom:1985ag}, where $V_{\mu\nu}$ and $B_{\mu\nu}$ are the field strengths of the new vector 
particle $V$ and the SM hypercharge, respectively. After diagonalizing the kinetic term, $V$ acquires a small charge to fields carrying hypercharge. Since the RHNs, $N$, do not carry hypercharge, $V$ only couples to $N$ via their mixing with LH neutrinos; the production rate of $N$ is consequently very small. 

As an alternative to kinetic mixing, the new gauge boson $V$ may couple directly to SM fields, which must carry a charge under the new $\mathrm{U}(1)'$. Suggestively, the SM is invariant under an accidental \emph{global} $\mathrm{U}(1)$ symmetry, namely baryon number minus lepton number ($B-L$). If this symmetry is instead a \emph{local} symmetry, the gauge theory suffers an anomaly in the $\mathrm{U}(1)_{B-L}^3$ triangle diagram; the theory is only consistent with three additional RHNs. Thus, RHNs are motivated by and naturally accompany gauge extensions of the SM. In general, there are other possible gauge symmetries that are combinations of baryon number and lepton flavour  and are also anomaly-free~\cite{He:1990pn,He:1991qd}.
The least constrained example in this category is $L_\mu - L_\tau$ symmetry, which still admits a 
``stronger-than-weak" strength of the new $U(1)'$ force \cite{Altmannshofer:2014cfa,Altmannshofer:2014pba}. However, in this model
$N$ may or may not be charged under the $U(1)'$, which introduces an extra degree of uncertainty on the presence and couplings of $N$, and we choose instead to concentrate solely on $B-L$.

How could Majorana RHNs coexist with this new gauge symmetry?
Given the strong constraints on new long-range forces, it is reasonable to expect that the new gauge boson is massive, which can be realized via the Higgs mechanism as in the SM. Then, the same scalar field that gives mass to the vector $V$ can also generate a Majorana mass for the RHN, thus tying $M_N$ to the scale of symmetry breaking and $M_V$. 
For example, if the breaking of the $\mathrm{U}(1)_{B-L}$ symmetry occurs due to the condensation of a scalar field $\Phi$  with  charge  $-2$ under $\mathrm{U}(1)_{B-L}$, then a Yukawa interaction of the form $y_N \Phi NN/2 + \mathrm{h.c.}$ will induce a Majorana mass for $N$
that is fully consistent with the gauge symmetry. Moreover, the spontaneous breaking of $\mathrm{U}(1)_{B-L}$ leads to masses for both $V$ and $N$, thus 
implying the relation 
\be
\frac{M_N}{M_V} \sim \frac{y_N}{g'}.
\ee
The lightness of $V$ would imply the lightness of $N$ if the gauge couplings and Yukawa couplings are of the same order. 
Thus, a $B-L$ gauge symmetry can be consistent not only with Dirac neutrino masses, but also with heavy Majorana neutrinos potentially in the same mass range as $M_V$.

\subsection{A Simplified Model}

With a $\mathrm{U}(1)_{B-L}$ gauge symmetry, the SM must be supplemented with three RHNs charged under the symmetry. Furthermore, to account for the observed LH neutrino mass splittings and mixing angles, there must be at least two RHNs with non-zero Yukawa couplings to the lepton doublet fields; this results in many parameters for the model that obscure the relevant phenomenology in high-energy experiments. We therefore  investigate a  simplified model with only one species of RHN, and this $N$  mixes with only one flavor of SM neutrino (namely, $\nu_\mu$)\footnote{For a detailed study of neutrino mixing parameters and RHN lifetimes in a full three-neutrino model, along with the phenomenology of prompt $N$ decays, see for example Ref.~\cite{Perez:2009mu}.}. This gives a more limited parameter space that can be thoroughly studied and facilitates comparison with other experimental tests of RHNs (see, for example, Refs.~\cite{Atre:2009rg,Deppisch:2015qwa,Das:2015toa}).  We emphasize, however, that a broader range of signatures is possible in the full model with several mixing angles, and experimental studies should be devised so as not to exclude sensitivity to, for instance, $N$ mixing with multiple flavors of lepton.

After the breaking of electroweak symmetry and the $\mathrm{U}(1)_{B-L}$, the RHN acquires a Majorana mass and mixes with the LH neutrino according to Eq.~(\ref{eq:theta_def}). The sterile state $N$ acquires a small charge under the electroweak gauge interactions through this mixing. We assume that the uneaten component of the $\Phi$ field responsible for breaking $\mathrm{U}(1)_{B-L}$ is heavy and decouples from the spectrum. Using two-component Weyl spinors, we write the Lagrangian of the model as:
\bea
{\cal L} = {\cal L}_{\rm SM} - \frac{1}{4}V_{\mu\nu}^2 - \frac{1}{2} M_V^2 V_\mu^2+ i N^\dagger \bar\sigma^\mu\partial_\mu N ~~~~~~~~~
\nonumber\\- \frac{M_N}{2}( N^2 + \mathrm{h.c.} )
+ g'V_\mu \left( \sum_{\rm SM} Q_{B-L}\psi^\dagger \bar\sigma^\mu \psi +N^\dagger \bar\sigma^\mu N  \right) 
\label{basis}
\\\nonumber
+ \theta_{\mu N} ~ \frac{g_W}{\sqrt{2}}\left(  \mu_L^\dagger \bar\sigma^\mu W_\mu^- N    + \mathrm{h.c.} \right) + \dots,~~~~~~~
\eea
as well as additional couplings of $N$ to $\nu$ and the $Z$/Higgs boson (analogous to the $W$ coupling) that we do not show explicitly here. 
SM lepton (antilepton) fields have charges $-1$ ($+1$), SM quark (antiquark) fields have charges $+1/3$ ($-1/3$), and the RHN fields have charge $+1$ to cancel the $\mathrm{U}(1)^3_{B-L}$ gauge anomaly.

The model has four unknown parameters:~$M_V$, $g'$, $M_N$, and $\theta_{\mu N}$. Our main goal is to investigate whether signals of $pp\rightarrow V\rightarrow NN$ in
existing and planned experiments will achieve sensitivity to $\theta_{\mu N}$ down to $\theta_{\rm s-s}$ given by Eq.~(\ref{thetass}), and if this  $B-L$ parameter space is currently allowed by all other experiments. In the following sections, we review the production and decay modes of both $V$ and $N$, and then discuss the current constraints on each.

\subsection{Production and decay of $V$ and $N$}

\noindent {\bf Gauge boson:}~There are several well-established production channels for $V$. These include meson decays, nucleon bremmstrahlung and direct 
Quantum Chromodynamics (QCD) production, as discussed in a recent review \cite{Alekhin:2015byh}. For the latter, the dominant channel is $ q \bar q \to V$ (as shown in Fig.~\ref{fig:N_pair_prod}) and $qg \to Vq$. 
For the LHC energies only the QCD production is relevant, while for SHiP all three production channels may be important. 
Light vector masses $M_V \sim1$~GeV and below can be considered as a dividing point below which the forward production of $V$ cannot be 
treated using the perturbative QCD approach. For this paper, we conservatively concentrate on the QCD production, and restrict our study to $M_V \gsim 1$~GeV, while noting that 
forward production for smaller masses would require an approach involving hadronic models. 

The most favorable spectrum for RHN pair-production is $M_V > 2 M_N$, in  which case on-shell $V$ bosons produced in the primary collisions subsequently decay to two $N$ particles. 
The partial decay width for $V\rightarrow NN$ is given by
\be
\Gamma_{V\to N N} = \frac16 \frac{(g')^{2}}{4\pi} M_V\left(1-\frac{4M_N^2}{M_V^2}\right)^{3/2},
\ee
while the decay rate of $V$ to (approximately massless) charged leptons, quarks, and neutrinos are given by 
\be
\Gamma_{V \to \ell\bar \ell} = 2\Gamma_{V \to \nu\bar\nu} =3 \Gamma_{V \to q\bar q}=\frac13 \frac{(g')^{2}}{4\pi} M_V.
\ee
Using these formulae, it is easy to see that the branching ratio of a GeV-scale $V$ boson to a pair of $N$ fermions is  of $\mathcal{O}(10\%)$. \\

\noindent {\bf Right-handed neutrino:}~The dominant production mode we consider for $N$ is  the pair production mode $V\rightarrow NN$ as shown in Fig.~\ref{fig:N_pair_prod}. The decays of $N$, however, proceed through its couplings to electroweak gauge and Higgs bosons (see Fig.~\ref{fig:N_decay}):~the couplings of $N$ are identical to the couplings of $\nu_\mu$ times the multiplicative factor $\theta_{\mu N}$. $N$ can therefore decay via $N\rightarrow {W^\pm}^{(*)}\mu^\mp$, $N\rightarrow Z^{(*)}\nu_\mu$, and $N\rightarrow h^{(*)}\nu_\mu$. The decay of $N$ depends crucially on its mass. For illustrative purposes, we show the leptonic 
decay rate, which in the limit $M_N\ll M_W$ is
\bea\label{eq:leptonicdecay}
\Gamma_{N \to \mu \ell_\alpha \nu_\alpha} &=& \frac{G_{\rm F}^2 M_N^5 |\theta_{\mu N}|^2}{192 \pi^3}\,\,\,\,\,\,\,(\alpha\neq\mu), \nonumber \\
\Gamma_{N \to \mu \mu \nu_\mu}&=& \frac{G_{\rm F}^2 M_N^5 |\theta_{\mu N}|^2}{192 \pi^3}(1+4\mathrm{s_W}^2+8\mathrm{s_W}^4), ~~~~
\eea
where $\mathrm{s_W} = \sin\theta_{\rm W}$ is the weak mixing angle and $G_{\rm F}$ is the Fermi constant. 
For $M_N\gtrsim1$ GeV, the hadronic decay width has a similar structure, although with additional color factors and quark mixing angle insertions. 
The scaling of the decay rate with the mass can be understood by substituting  $\theta_{\mu N} = \theta_{\rm s-s}$ from Eq.~(\ref{thetass}), 
\be
\Gamma_{N \to \mu \ell \nu} \simeq 10^{-15}~{\rm eV} \times \frac{|\theta_{\mu N}|^2}{\theta_{\rm s-s}^2} \left(  \frac{M_N}{1~{\rm GeV}}   \right)^4.
\ee
While $\Gamma_N$ scales like $M_N^5$ for fixed mixing angle, the mixing angle predicted by the see-saw relation also scales as $M_N^{1/2}$, leading to the fourth power scaling shown here. 
We see, therefore, that the decay width is very small for $M_N\ll M_W$ and exhibits a very strong power-law dependence on $N$. For $M_N\gtrsim M_W$, the two-body decay modes open and the width scales linearly with $M_N$ above this value.
Exclusive hadronic decay rates of $N$ relevant for very low masses can be found in \cite{Gorbunov:2007ak}.

\begin{figure}[t]
\centering
\includegraphics[width=0.48 \textwidth ]{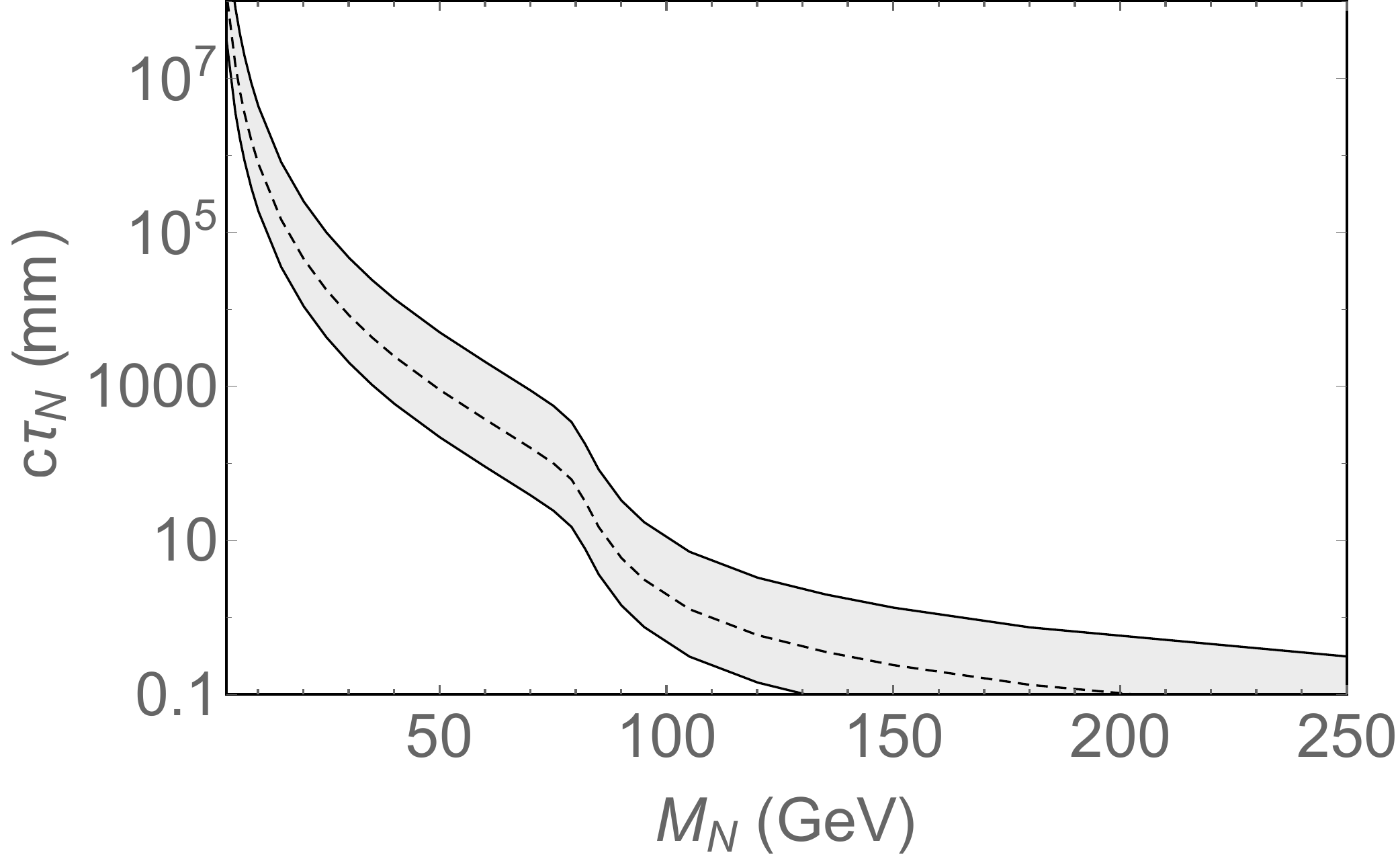}
\caption{Proper $N$ decay distance as a function of the RHN mass. In computing the lifetime, the mixing angle is fixed by using the single-neutrino see-saw relation, Eq.~(\ref{eq:mixing}), for various LH neutrino masses. The curves shown are: $m_\nu = \sqrt{\Delta (m_\nu^2)^{\rm sol}}$ (upper solid); $m_\nu = \sqrt{|\Delta (m_\nu^2)^{\rm atm}|}$ (middle dashed); $m_\nu=0.23$ eV (lower solid), which is equal to the current Planck limit on the sum of the neutrino masses \cite{Ade:2015xua}.
}
\label{fig:mixing_lifetime}
\end{figure}

Of particular relevance for us is that, for $M_N$ accessible at experiments such as SHiP and the LHC, the width is sufficiently small that the decay of $N$ typically occurs on macroscopic scales for mixing angles $\theta_{\rm s-s}$. We show the proper decay distance, $c\tau_N$, as a function of $M_N$ for various mixing angles motivated by the see-saw mechanism in Fig.~\ref{fig:mixing_lifetime}; we include all decay modes in this plot, not just those shown in Eq.~(\ref{eq:leptonicdecay}).

\subsection{Existing Constraints on $N$}

Most searches for RHNs do not assume any production modes beyond their mixing with LH neutrinos. There are several types of such direct search strategies for RHNs. The most relevant constraints on RHNs for the regions of parameter space relevant to us are shown in Figs.~\ref{fig:master_MN_V2_SHiP}-\ref{fig:master_MN_V2_LHC} \cite{Cooper-Sarkar:1985nh,Bernardi:1985ny,Abreu:1996pa,Vaitaitis:1999wq,Atre:2009rg,Gorbunov:2007ak,Boyarsky:2009ix,Ruchayskiy:2012si,Liventsev:2013zz,Deppisch:2015qwa}. They include:

\begin{enumerate} 

\item Searches for rare 
meson decays,  such as $K^\pm\to \mu^\pm + N$ (see, {\em e.g.} \cite{PIENU:2011aa,Artamonov:2014urb}), via a modification of  
the momentum spectrum of the charged lepton. The rate for such processes scales as $|\theta_{\mu N}|^2$. 

\item Searches for $N$ in beam-dump experiments (see, {\em e.g.,}~\cite{Bernardi:1987ek,Vaitaitis:1999wq,Alekhin:2015byh}) via production of GeV-scale $N$ in the rare decays of bottom and charm quarks ($b\to cl^- N$, $c\to sl^+ N$) or kaons ($K^\pm\rightarrow \mu^\pm N$), with subsequent visible decays of $N$ in a detector at some distance from the production target. Due to the decay length of $N$ exceeding the target-detector separation distance, the signal in such searches scales as the fourth power of mixing angle, $|\theta_{\mu N}|^4$, for proper decay lengths much longer than the distance from the dump to the detector. 

\item Finally, the relatively high-energy collider experiments at BaBar, Belle, LEP, and the LHC are sensitive to the 
production of both light $N$ (in meson decay) and heavier $N$, via prompt and displaced vertex searches \cite{Adriani:1992pq,Abreu:1996pa,Liventsev:2013zz,Khachatryan:2015gha}.
If decay occurs within the detector and can be triggered on, the sensitivity scales again as $|\theta_{\mu N}|^2$. 

\end{enumerate}

Cosmology also constrains the RHN scenario:~bounds from BBN strongly constrain see-saw mixing angles for $\mathrm{MeV}\lesssim M_N\lesssim400-1000$ MeV, depending on the precise mixing angle \cite{Gorbunov:2007ak,Boyarsky:2009ix,Ruchayskiy:2012si}. For $M_N \gsim 1000 $ MeV and $\theta \gsim \theta_{\rm s-s}$, there are no strong cosmological constraints as $N$ would decay within $\sim 0.1$ seconds. 

As an aside, the existence of the new vector portal for $N$ may {\em extend} 
the mass range for $M_N$ that is allowed by BBN. The range of masses, few~MeV $ \lsim M_N \lsim M_\pi$, which are excluded in the minimal model without new gauge interactions
by the arguments of $N$ stability during BBN neutron-proton freeze-out, may be allowed in the presence of $V$. 
If  $M_V < M_N$ the annihilation process $NN \to VV $ opens up, while for $M_V > M_N$ annihilation to visible neutrinos 
is important, $NN\to V^{*} \to \nu \nu$. The net effect will be the annihilation-driven depletion of  the cosmological abundance of $N$, with consequent weakening of the BBN bounds.

\subsection{Existing Constraints on $V$}

If $V$ is the gauge boson of a new $B-L$ force, we showed that it has a $\sim10\%$ branching fraction into $NN$ when kinematically allowed. However, this implies that 90\% of decays are into SM states, and so we expect strong constraints on the model from direct searches for $V$. We summarize these bounds in Fig.~\ref{fig:master_MV_gp}  \cite{Williams:2011qb,Bellini:2011rx,Chatrchyan:2013tia,Lees:2014xha,Khachatryan:2014fba,Aad:2014cka,Prasad:2015mxa}. There are several such searches:

\begin{enumerate}

\item $V$ induces elastic scattering between electrons and neutrinos that is constrained by the Borexino experiment \cite{Bellini:2011rx,Harnik:2012ni}. For $M_V$ well above the Borexino threshold of 200 keV, the constraint is approximately
\be
g' \lesssim 4\times10^{-3}\times\frac{M_V}{1\,\,\mathrm{GeV}}.
\ee

\item New gauge bosons can be produced via radiative return at electron-positron colliders, $e^+e^- \to \gamma V \to \gamma \mu^+  \mu^-$ \cite{Lees:2014xha}. For $M_V \gtrsim 1$ GeV up to the kinematic limit of $B$-factories, these constraints are stronger than from neutrino-electron scattering. LEP also constrains $V$ via the measurement of the hadronic cross section at $s=M_Z^2$ \cite{Williams:2011qb}.

\item $V$ contributes to Drell-Yan processes at hadron colliders, and stringent bounds exist on resonant contributions to $pp\rightarrow V\rightarrow \ell^+\ell^-$. The strongest limits come from the LHC. For masses $M_V< M_Z$, limits were estimated from the Drell-Yan spectrum measured by the CMS Collaboration at $\sqrt{s}=7$ TeV \cite{Chatrchyan:2013tia,Hoenig:2014dsa}, and extrapolated to 8 and 14 TeV (future colliders were considered in Ref.~\cite{Curtin:2014cca}). It should be emphasized, however, that these are estimates and the true limits may be somewhat weaker, particularly in the case of 14 TeV limits, which were  assumed optimistically to scale indefinitely with the square root of integrated luminosity. A recent proposal for a search at LHCb could have better sensitivity than ATLAS/CMS in the region $M_V\lesssim45$ GeV \cite{Ilten:2016tkc}. The Drell-Yan constraints  disappear for $M_V\sim M_Z$, as such regions are typically excluded from new resonance searches; a LEP-1 search by L3 for narrow quarkonium resonances in the vicinity of $M_Z$ was carried out and could yield slightly stronger constraints in this region than what we show, although it is not  apparent how to directly apply the L3 search to our model. 
With $M_V>M_Z$, constraints on $M_V$ production come from ATLAS and CMS measurements of the Drell-Yan spectrum above the $Z$ pole \cite{Khachatryan:2014fba,Aad:2014cka}.

\item New vector interactions can induce flavor-changing neutral currents in meson decays. The conservation of the $B-L$ current forbids these at tree level. 
Loop processes may lead to the $K^+ \to \pi^+ + V \to \pi^+ + \nu \bar\nu$ decays \cite{Pospelov:2008zw}, which will impose
some constraints on $g'$ if $M_V < M_K-M_\pi$. A conservative evaluation of this rate shows \cite{Batell:2014yra} that this constraint 
cannot compete with neutrino scattering. The same applies to the recent analysis of $\pi^0$ Dalitz decays \cite{Goudzovski:2014rwa}.

\end{enumerate}

Finally, the existence of a coupling between $N$ and $V$ can thermalize $N$ in the early universe.  If there is a very light RHN, it can be overabundant and lead to constraints from excess energy in radiation. The strongest constraints apply to the pure-Dirac case \cite{Heeck:2014zfa}, whereas we consider $N$ that are sufficiently heavy to have quickly decayed prior to BBN, and so these cosmological constraints are not applicable to our scenario.

%% file: LHC.tex
Since $B-L$ gauge bosons have an appreciable coupling to quarks, hadron colliders are ideal experiments for discovering a new $B-L$ gauge interaction. In this section, we argue for the importance of $pp\rightarrow V\rightarrow NN$ signatures, where the $N$ decays at a displaced vertex (DV).

 Conventionally, discovery of $V$ is easiest in the dilepton final state, $pp\rightarrow V\rightarrow\ell^+\ell^-$, due to the signal resonance, relatively  low SM backgrounds and high lepton-identification efficiencies. However, electroweak backgrounds are large for dilepton invariant masses $\lesssim$ few hundred GeV, and because of the finite invariant mass resolution of the detector, such searches are background-limited with sensitivity growing at best as the square root of integrated luminosity. Sensitivity may also be limited by uncertainties in background modeling or other effects at high luminosity. By contrast, the spectacular displaced  decays of $N$ can lead to final states with much lower SM backgrounds; indeed, some searches are expected to remain background free even throughout the high-luminosity phase of LHC running. In the regime where $N$ is long-lived and decays at a DV, as is true for much of the see-saw parameter space with $M_N\sim10-100$ GeV (see Fig.~\ref{fig:mixing_lifetime}), such searches can be background-free and so the sensitivity instead scales linearly with luminosity. Thus, at high luminosity the sensitivity for $pp\rightarrow V\rightarrow NN$ can be superior to that for dilepton resonances, and RHNs can serve as a discovery mode for $V$ with projected sensitivities  down to $g'\sim\mathcal{O}(10^{-4})$.

At least half of the RHNs produced at the LHC decay via the charged current interaction, and so most events have at least one displaced lepton and additional displaced hadrons and/or leptons. Because $N$ are produced in pairs, this gives a striking signature; most LHC Run 1 analyses are background-free requiring only a single DV in the inner detector (or two displaced leptons), and so it is expected that a background-free analysis for two DVs can be devised through the end of high-luminosity running while maintaining a reasonable signal efficiency. It should be noted, however, that the DV searches are most powerful relative to dilepton searches where the backgrounds for the competing dilepton search are largest, namely at low invariant masses for $V$; thus, dedicated searches may be necessary to keep reconstruction thresholds sufficiently low to efficiently tag one or two DVs from signal processes. This is in contrast with some DV searches motivated by supersymmetry, where new states have masses well above the weak scale and very stringent kinematic cuts can suppress backgrounds while maintaining high signal efficiency.

In this section, we  review the existing DV searches relevant for $pp\rightarrow V\rightarrow NN$ production at the LHC, most of which look for a single displaced object. We then project the sensitivity of the high-luminosity (HL) LHC  to the $B-L$ model parameter space, showing extrapolations of current searches as well as proposals for searches for two DVs that can retain sensitivity in case the backgrounds in the single displaced vertex analyses become unmanageably large.

\subsection{Overview of Current Displaced Vertex Searches}

In Run 1, ATLAS, CMS, and LHCb have each performed analyses sensitive to the decays of long-lived particles in various components of the detector. These searches range from very inclusive studies to highly optimized searches for particular models. Due to the limited acceptance and integrated luminosity of LHCb, we focus on searches in ATLAS and CMS, highlighting those most relevant for RH neutrino decays; however, recent studies have shown that LHCb could have good sensitivity to some models with low-mass vectors, and this is an interesting direction for follow-up studies \cite{Ilten:2016tkc}.

We now summarize the relevant searches at ATLAS and CMS. \\

\noindent {\bf Displaced dilepton search, no vertex requirement (CMS):}~CMS performed a search for ``displaced supersymmetry (SUSY)'' \cite{Khachatryan:2014mea}, sensitive to final states with two high-impact-parameter\footnote{The impact parameter is the point of closest approach of a track to the primary vertex when extrapolated back towards the collision point.}, opposite-flavor leptons. The search is agnostic about any other high-impact-parameter tracks in the event, and no DV is explicitly reconstructed. Events are selected with at exactly one electron and muon with $p_{\rm T}>25$ GeV and $|\eta|<2.5$ each. The leptons must be isolated from one another, from jets, and from other high-$p_{\rm T}$ tracks or energetic calorimeter depositions. For the signal region where both lepton transverse impact parameters ($|d_0|$) are between 1-20 mm, no events were observed with an expected background of approximately $0.05\pm0.02$ events. CMS tracking can be moderately efficient out to $|d_0|\sim20$ cm \cite{CMS:2014hka}, and so it is expected that the search could be extended to higher displacements without a substantial increase in background rate.\\

\noindent {\bf Displaced dilepton vertex search (CMS):}~There is a CMS search for DVs containing either two electrons or two muons \cite{CMS:2014hka}. The leading electron must have $E_{\rm T}>40$ GeV, with other leptons satisfying $p_{\rm T}\gtrsim25$ GeV. The leptons must be isolated from other high-$p_{\rm T}$ tracks, but not from one another. The two leptons must reconstruct a DV, have large impact parameter significant (roughly equivalent to a requirement $|d_0|\gtrsim0.2$ mm), satisfy $M_{\ell\ell}>15$ GeV, and the dilepton vector must point within the same azimuthal semicircle as the line from the primary vertex to the DV. Cosmic ray muons are suppressed by vetoing back-to-back muons. Zero  events are observed, with the expected background not quantified but expected to be much less than one; indeed, no events are observed even in the control region. \\

\noindent {\bf Displaced lepton + hadrons vertex search (ATLAS):}~This is an ATLAS search for a DV containing muons plus tracks \cite{Aad:2015rba}. The event is triggered by a muon with $p_{\rm T}>50$ GeV, or an electron with $E_{\rm T}>120$ GeV (or two electrons with $E_{\rm T}>40$ GeV each; the electrons are selected using photon triggers that do not require a track). DVs are selected by reconstructing tracks with transverse impact parameter $|d_0|>2$ mm, and transverse vertex displacements must be larger than 4 mm. Vertices with five or more tracks, a track invariant mass $>10$ GeV, and containing at least one lepton are selected; in the muon+tracks channel of most relevance to our analysis, the estimated background is $\sim10^{-3}$. It should be noted that no isolation requirements are applied to the leptons. No events are observed with $\ge5$ tracks, even for vertex masses below 10 GeV, suggesting that relaxing the mass requirement somewhat (while potentially introducing isolation cuts) should not introduce appreciable backgrounds and could improve sensitivity to lower-mass displaced long-lived objects.\\

\noindent {\bf Displaced dilepton vertex search (ATLAS):}~ATLAS has also searched for pairs of leptons from a single DV  \cite{Aad:2015rba}. The trigger requirements are the same as for the displaced lepton + hadrons vertex search described above. Each lepton must have $p_{\rm T}>10$ GeV and $|d_0|>2$ mm, and cosmic ray muons are suppressed by vetoing back-to-back muons. No isolation requirements are applied to the leptons, and the invariant mass of all tracks at the vertex must exceed 10 GeV. No dilepton vertices are observed in the signal region, and only a few are observed even for $M_{\ell\ell}<10$ GeV, with a background estimate in the signal region  of $\mathcal{O}(10^{-3})$ events. \\

\noindent {\bf DVs in muon spectrometer (ATLAS):}~There is an ATLAS search for pairs of hadronic DVs in the muon spectrometer (MS)\footnote{The analysis also looks for a vertex in the MS coincident with a DV in the inner detector; however, the signal rate is typically higher for both particles to decay in the MS when the $N$ decay length is long enough to reach the MS, so we focus on this case.} \cite{Aad:2015uaa}. The analysis relies on a dedicated trigger  sensitive to clusters of activity in the MS without corresponding energy depositions in the inner detector or calorimeters. This trigger is sensitive to low-mass, long-lived particles whose traces may not be energetic enough to otherwise allow the event to be recorded. However, the probability of having two long-lived particles decaying in the sensitive regions of the MS is small, which hurts signal sensitivity. The analysis observes only two background events.  

\subsection{Recasts of Current Searches}\label{sec:recast}

\begin{figure}[t]
\centering
\includegraphics[width=0.5 \textwidth ]{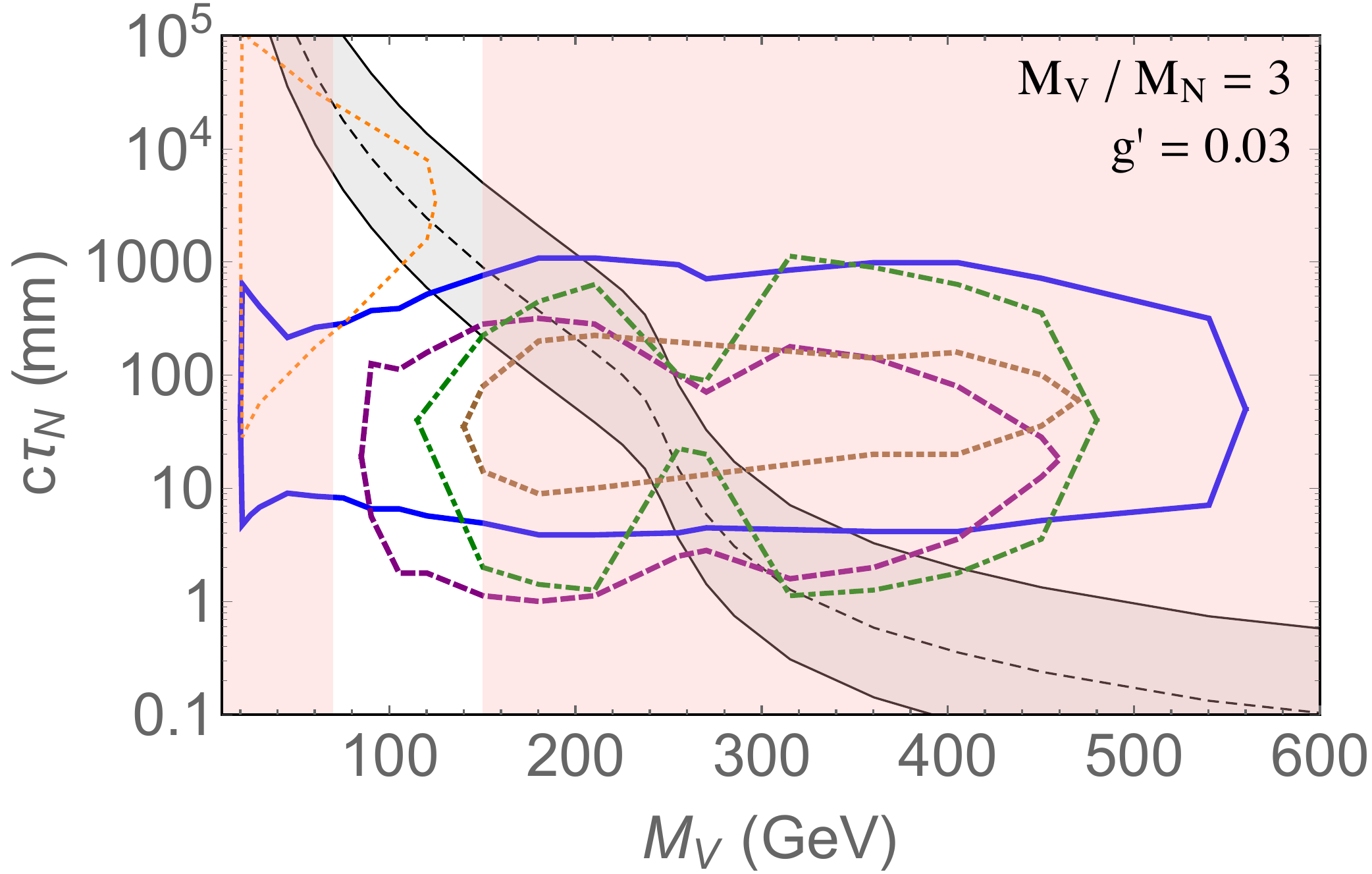}
\caption{Projected constraints on a new $\mathrm{U}(1)_{B-L}$ gauge boson from LHC Run 1 searches for the displaced decay of $N$ ($M_V/M_N=3$ and $g'=0.03$). Searches considered are the ATLAS displaced dilepton vertex search (blue, solid) \cite{Aad:2015rba}; ATLAS displaced muon + tracks vertex search (brown, dotted) \cite{Aad:2015rba}; CMS displaced dilepton vertex search (green, dot-dashed) \cite{CMS:2014hka}; CMS displaced dilepton search without vertex requirement (purple, dashed) \cite{Khachatryan:2014mea}; ATLAS muon spectrometer vertex search (orange, thin dotted) \cite{Aad:2015uaa}.  The grey shaded region shows the preferred parameter space for obtaining the LH neutrino masses from Fig.~\ref{fig:mixing_lifetime}. Shaded red regions are excluded from CMS \cite{Chatrchyan:2013tia,Khachatryan:2014fba} and ATLAS \cite{Aad:2014cka} dilepton resonance searches for $pp\rightarrow V\rightarrow \ell^+\ell^-$.   }
\label{fig:MV_ctau_run1}
\end{figure}

\begin{figure}[t]
\centering
\includegraphics[width=0.5 \textwidth ]{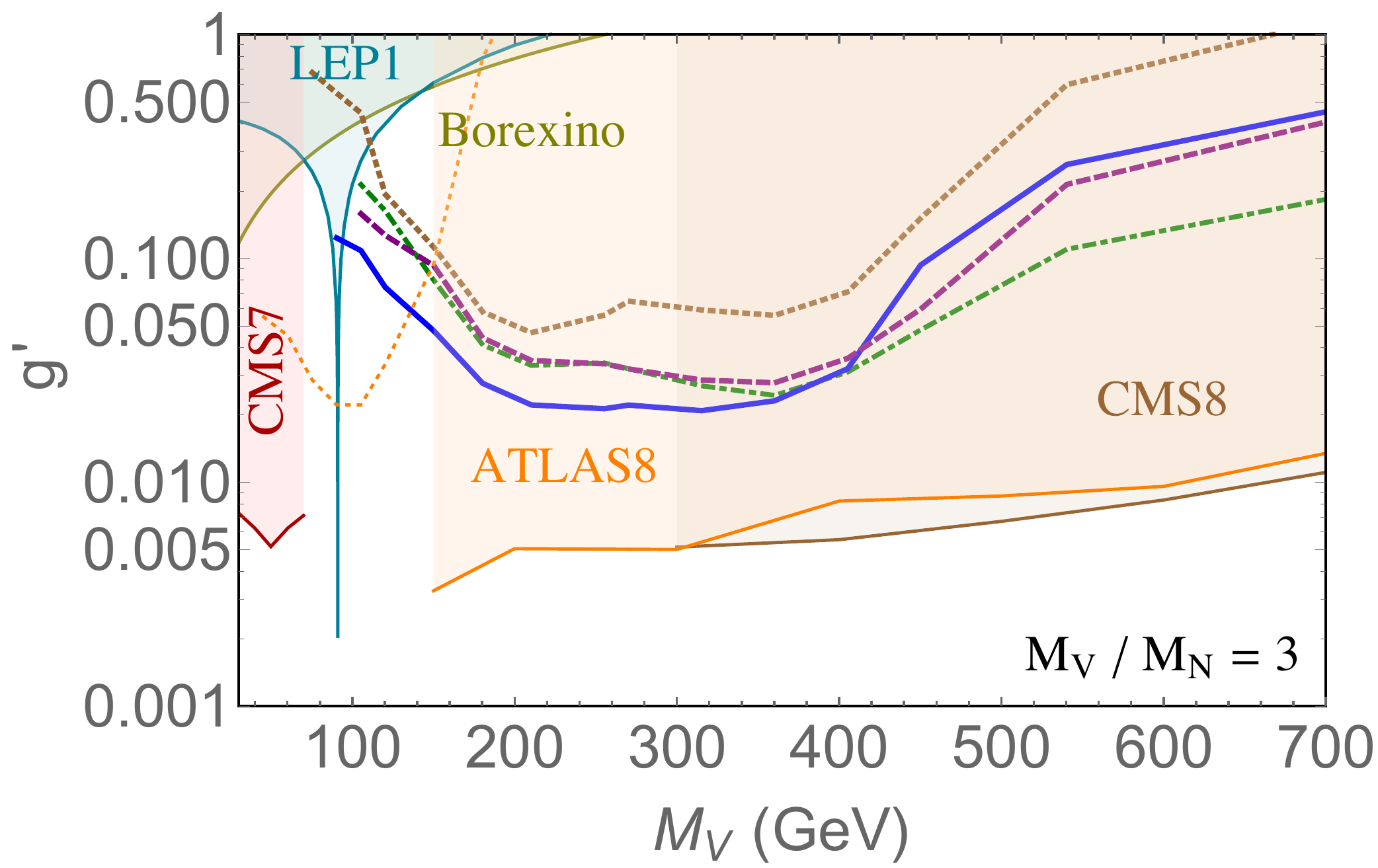}
\caption{Projected constraints on the mass and gauge coupling for a new $\mathrm{U}(1)_{B-L}$ gauge boson from LHC Run 1 searches for the displaced decay of $N$ ($M_V/M_N=3$). The displaced vertex search projections are the same as in Fig.~\ref{fig:MV_ctau_run1}, while the other bounds on the gauge boson were described in Sec.~\ref{sec:simplified_model}. The RH neutrino mixing angle is fixed using Eq.~(\ref{thetass}). }
\label{fig:MV_gp_run1_matm}
\end{figure}

\begin{figure}[t]
\centering
\includegraphics[width=0.5 \textwidth ]{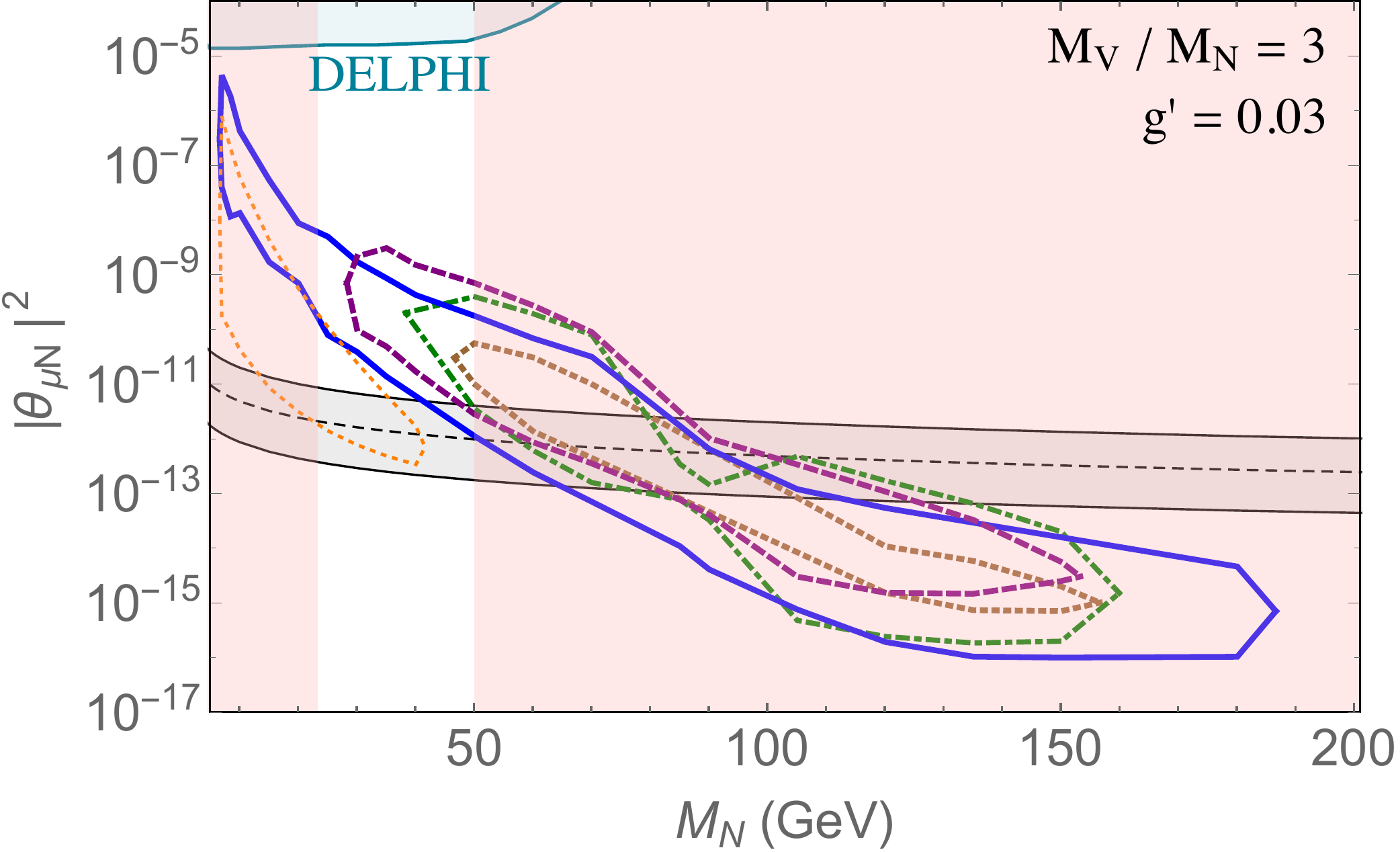}
\caption{Projected constraints on the RHN mass and mixing angle in a model with gauged $\mathrm{U}(1)_{B-L}$ ($M_V/M_N=3$ and $g'=0.03$). The displaced vertex search projections are the same as in Fig.~\ref{fig:MV_ctau_run1}, as is the shaded red region from dilepton searches for $V$. Other bounds on the RHN parameter space were described in Sec.~\ref{sec:simplified_model}.  }
\label{fig:MN_V2_run1_gp3em2}
\end{figure}

Our estimates for constraints of Run 1 DV searches on the RHN-$\mathrm{U}(1)_{B-L}$ parameter space are shown in Figs.~\ref{fig:MV_ctau_run1}-\ref{fig:MN_V2_run1_gp3em2}. As expected, the DV searches are sensitive to parameters that explain the observed neutrino masses. In particular, Fig.~\ref{fig:MN_V2_run1_gp3em2} shows that $V\rightarrow NN$ searches can probe RHN mixing angles many orders of magnitude below direct searches for $N$.  However,  dilepton searches for $pp\rightarrow V\rightarrow \ell^+\ell^-$ are currently more powerful than the DV searches for $V\rightarrow NN$. The main exception to this statement is that dilepton resonance searches are typically insensitive to $M_V\sim M_Z$ because such masses are excluded from the signal regions of the corresponding analyses. We expect, however, that the different scaling of the sensitivity for the background-dominated dilepton searches vs.~the background-free DV searches will result in the DV searches being  more powerful at the HL-LHC.

We now describe the methods of our recasts in more detail. None of the existing DV searches consider long-lived RHNs as a benchmark model. The  efficiency of reconstructing DVs depends on many different properties of a signal, such as the kinematics of the final-state particles, the opening angle between tracks, and the location of decay in the detector. It is not possible to correctly include these effects without a full-scale detector simulation and validation; however, the experimental analyses typically provide some efficiency information for other benchmark models that can be extrapolated to estimate the efficiencies for DVs from RHN decay. Thus, we can estimate the approximate sensitivity  to RH neutrinos of current DV searches, but the precise bounds depend on model-dependent efficiencies that must be determined by the experimental collaborations.

For this and subsequent analyses, we used a UFO model developed using the \texttt{FeynRules} package \cite{Degrande:2011ua,Alloul:2013bka}. Signal events of $pp\rightarrow V\rightarrow NN$ were generated using \texttt{MadGraph5\_aMC@NLO} \cite{Alwall:2014hca}, and $N$ were subsequently decayed using the \texttt{MadSpin} package \cite{Frixione:2007zp,Artoisenet:2012st}. Parton-level events were generated with up to one additional final-state parton and showered with \texttt{Pythia 6} \cite{Sjostrand:2006za}; parton-level events of different multiplicity were merged with the shower using the MLM-based shower-$k_\perp$ scheme \cite{Alwall:2008qv}. 

In recasting existing analyses, we first reconstruct all leptons, tracks, and vertices at truth level. We then apply efficiencies for lepton, displaced track, and DV reconstruction according to the efficiencies given in a specific analysis. As mentioned earlier, the kinematics of our signal are rarely identical to one of the signal benchmarks in a given analysis; we therefore select the efficiencies for the signal benchmark that most closely reproduces the kinematics of our $V\rightarrow NN$ signal. A comparison of  efficiencies between various  benchmark models provided in each ATLAS analysis suggests that our estimated Run 1 cross section limits should be correct to within better than a factor of two, even though we do not have the exact efficiency information\footnote{Specifically, we use the efficiencies from the following benchmark models:~hidden valley, $m_{\pi_{\rm V}}=25$ GeV for the ATLAS MS vertex search \cite{Aad:2015uaa}; $m_{\tilde q}=700$ GeV, $m_{\tilde\chi^0}=108$ GeV for the ATLAS muon + tracks search \cite{Aad:2015rba}; $m_{\tilde g}=600$ GeV, $m_{\tilde\chi^0}=400$ GeV for the ATLAS displaced dilepton search  \cite{Aad:2015rba}; we use a flat efficiency-to-acceptance ratio of 35\% for the CMS displaced dilepton vertex search (as discussed in Section 4 of Ref.~\cite{CMS:2014hka}); for the CMS displaced SUSY search \cite{Khachatryan:2014mea}, we use standard lepton identification efficiencies multiplied by a $|d_0|-$dependent track efficiency \cite{CMS:2014hka}.}. One of the most significant factors that hinders reconstruction of a DV is the boost of the parent particle, since boosted decays give collimated sprays of particles that point back towards the primary vertex \cite{Aad:2015rba}; consequently, boosted long-lived decays are more likely to be mis-modeled by a simplistic theorists' analysis \cite{Cui:2014twa,Liu:2015bma}. We therefore choose a benchmark scenario for which the $N$ are relatively unboosted: $M_V/M_N=3$.

Our projected reach for Run 1 searches is shown in Fig.~\ref{fig:MV_ctau_run1} in the $M_V-c\tau_N$ plane for a particular value of $g'=0.03$. As expected, the searches for DVs in the inner detector are sensitive to $c\tau_N\sim1\,\,\mathrm{mm}-1\,\,\mathrm{m}$, while searches for vertices in the MS are sensitive to proper decay lengths in the $1-10$ m range. The Run 1 DV searches are also sensitive to RHN lifetimes motivated by the see-saw mechanism as illustrated by the grey shaded region in Fig.~\ref{fig:MV_ctau_run1}. However, the constraints from  $V\rightarrow \ell^+\ell^-$ are currently stronger than the DV limits, and the DV searches have no sensitivity for $g'$ below the dilepton bounds where they exist. The principal exception is for $M_V\approx M_Z$ due to the complications of a resonance search in the vicinity of the $Z$ pole; this region is typically used to normalize the dilepton spectrum and is therefore excluded from searches for dilepton resonances. The DV searches, however, have no restriction in covering masses around the $Z$, and currently offer the best limits for this mass range. 

We remark further on one peculiar feature in the DV sensitivity curves for $M_V\approx240$ GeV:~here, $M_N \approx M_W$, and so the two-body decay $N\rightarrow W^\pm \mu^\mp$ begins to dominate. Since the two-body decay is close to threshold, the muon is very soft and there is a sharp decline in sensitivity immediately around this mass; for higher masses, the muon is once again sufficiently energetic to pass the trigger and reconstruction requirements of the DV searches.

Further estimates for the Run 1 DV sensitivity to the RHN-$\mathrm{U}(1)_{B-L}$ parameter space are shown in Figs.~\ref{fig:MV_gp_run1_matm}-\ref{fig:MN_V2_run1_gp3em2}. It is clear that DV searches would be sensitive to the neutrino mass parameter space motivated by the minimal see-saw mechanism and would be well below the reach of other searches for direct RHN production; however, dilepton constraints currently already exclude these values of $g'$. The performance of DV searches is also suboptimal because the analyses are not  configured for the RHN signal:~either they require opposite-flavor leptons (unlike our simplified model, which predominantly gives same-flavor leptons), require dilepton vertices (whose rates are suppressed by leptonic $W/Z$ branching fractions), or have high thresholds and low reconstruction efficiencies. With dedicated searches and increased integrated luminosity, the lack of backgrounds in the DV searches make them very important probes of RHNs in LHC Run 2 and beyond.

\subsection{Prospects for Future LHC Running}

Although Run 1 DV searches are typically not the most powerful probes of the $\mathrm{U}(1)_{B-L}$ model, the fact that DV searches are background-free and may remain so throughout high-luminosity running means that their sensitivity relative to $V\rightarrow\ell^+\ell^-$ constraints grows linearly with integrated luminosity. Indeed, DV searches are one of the rare examples in which the sensitivity to new physics production cross section remains linear throughout high-luminosity running, provided that trigger thresholds can be kept low and vertex reconstruction is not overly hindered by the high pile-up conditions.

In this section, we quantify the expected sensitivity of DV searches to $V\rightarrow NN$ after high-luminosity running (HL-LHC:~$3\,\,\mathrm{ab}^{-1}$ of integrated luminosity at $\sqrt{s}=14$ TeV). In order to determine the HL-LHC reach, we must estimate the backgrounds; this can only be done by extrapolating the current Run 1 analyses. Assuming a linear dependence of background events on the luminosity, the inner-detector DV searches described in Sec.~\ref{sec:recast} predict $\mathcal{O}(\mathrm{few})$ background events, while the MS DV search predicts $\mathcal{O}(100-1000)$ events. However, there are a number of factors that can affect this prediction:~the very high pile-up encountered in the HL-LHC could degrade vertex reconstruction and also give more accidental track crossings at high displacement, resulting in a higher background than na\"ively predicted. Conversely, the ATLAS and CMS detectors will be upgraded to cope with the larger number of primary vertices, and these new capabilities could improve background rejection. Improvements to the  algorithms for vertex tagging and high-impact-parameter track reconstruction could give still further gains.

Because of this uncertainty, we provide projections of signal sensitivity for two different scenarios. In the first, we propose a search for pairs of displaced objects in the inner detector which should remain background-free even in very high pile-up conditions. Second, we show the results from an extrapolation of current Run 1 searches through HL running. For searches with vertex reconstruction in the MS, we only show results that are extrapolations of current searches due to the challenges of modeling vertex reconstruction in the MS. \\

\noindent {\bf Inner Detector DV Searches:}~The current Run 1 searches are background free when requiring a DV with a lepton + hadrons, or two displaced leptons (without necessarily reconstructing a vertex). At the HL-LHC, these may no longer be background free although the backgrounds are expected to be very small. Given the rarity of finding one of these signals in Run 1 data, the \emph{combination} of two should remain background-free throughout HL running even with very high pile-up conditions\footnote{For example, the expected background cross section for the CMS search in Signal Region 3 for two displaced leptons (without vertex) is $\sim$ab. Assuming the leptons are uncorrelated, this gives a mistag probability for a single displaced, isolated lepton of $\lesssim10^{-6}$, which is more than enough to suppress  backgrounds associated additional displaced objects. Similar arguments are presented in the Appendix of Ref.~\cite{Cui:2014twa}.}. This allows us to remove the uncertainty in background estimation from our projections, and we show signal sensitivity to \emph{\bf five events} with $3\,\,\mathrm{ab}^{-1}$. 

In fact, the background suppression of an additional displaced object beyond the Run 1 searches should allow for the relaxation of other requirements such as DV selection criteria or kinematic thresholds. Given the potentially very small signal rates, maximizing signal efficiency is of utmost importance:~it is important to consider the possible gains of relaxing DV selection criteria vs.~the inefficiency of having to select additional displaced objects. 

\emph{Trigger:}~Triggering is a major challenge for the HL-LHC, since lepton trigger thresholds must be kept low to retain sensitivity to leptonic Higgs decays and other electroweak final states. This will likely necessitate the use of tracking information at trigger Level 1 (L1) as well as at higher levels. In the case of DV signals, this can be both beneficial and harmful:~trigger requirements that require an association of leptons with prompt tracks would make it more challenging to trigger on displaced leptons as in the RHN model, whereas the availability of tracking information at lower levels of the trigger could allow for the selection of events with many displaced tracks (or, alternatively, many ``trackless'' objects), allowing for  lower thresholds. It is impossible to say with certainty what the trigger capabilities and limitations of ATLAS and CMS will be in HL running, and so we consider a trigger scenario  consistent with some of the projections for L1 thresholds at the HL-LHC and/or current lepton triggers (for example, see Ref.~\cite{Butler:2020886}):
\bi
\item Single isolated lepton with $p_{\rm T}>25$ GeV, OR
\item Two isolated leptons, each with $p_{\rm T}>15$ GeV, OR
\item Three muons, each with $p_{\rm T}>6$ GeV.
\ei
The triggers for electrons will likely be higher, but since the simplified model under consideration gives muon-rich sinagures, this suffices for our analysis. For comparison, we also show results for a more pessimistic menu with higher thresholds:~$p_{\rm T}>35$ GeV for single muons (45 GeV for electrons); $p_{\rm T}>25$ GeV for muons in the dilepton trigger (30 GeV for electrons); and $p_{\rm T}>10$ GeV for the 3-muon trigger.

\emph{Event selection:}~We select events where one $N$ decays semileptonically (\emph{i.e.}, $N\rightarrow \mu^\pm q\bar q'$), and the other decays to at least one lepton. This gives rise to a distinctive signature of one DV with a muon + several hadronic tracks, and there is an additional displaced lepton unassociated with the vertex. This is inspired by a combination of the CMS ``displaced supersymmetry'' \cite{Khachatryan:2014mea} analysis with the ATLAS muon + tracks analysis \cite{Aad:2015rba}. The leptons are required to be isolated from hadronic activity and a flat identification efficiency of 90\% (70\%) is applied for muons (electrons). The leptons considered in the analysis must have $p_{\rm T}>5$ GeV (10 GeV) for muons (electrons), although the leptons are typically harder than this in order to pass the dilepton triggers. We require that the event have a DV containing a muon and at least four other tracks with $p_{\rm T}>1$ GeV; the total invariant mass of the tracks must exceed 6 GeV to suppress heavy-flavor backgrounds. Back-to-back muons are vetoed to suppress cosmic ray backgrounds.

In reconstructing displaced objects, we require displaced tracks to have a transverse impact parameter $1\,\,\mathrm{mm} < |d_0| < 30\,\,\mathrm{cm}$, and we apply a $|d_0|$-dependent reconstruction efficiency for each track \cite{CMS:2014hka}. We refrain from using DV tagging efficiencies from specific current searches because we wish to consider the possibility of searches that deviate from the current benchmarks for vertex tagging. We  require that tracks originate within 60 cm of the primary vertex in the radial direction ($r_0$) and 50 cm in the longitudinal direction ($z_0$). Because this method has been shown to over-estimate the vertex reconstruction efficiencies in some current searches \cite{Liu:2015bma}, we also show results for a more pessimistic tagging scenario based on approximate DV tagging efficiencies derived in Ref.~\cite{Liu:2015bma} that appear to replicate current DV searches with reasonable accuracy. In the pessimistic case, we apply additional efficiencies that penalize the reconstruction of tracks that originate close to the edge of the tracking system:~these are linearly falling functions of $|d_0|$, $r_0$, and $|z_0|$ that are fully efficient at the primary vertex and zero at the edge of the allowed region. We also apply an additional reconstruction efficiency for each vertex that falls quadratically in $|d_0|$ from fully efficient at the origin to zero at $|d_0| = 30$ cm.

\emph{Results:}~We employ the same MC simulation strategy  described in Section \ref{sec:recast}, with events generated at $\sqrt{s}=14$ TeV and assuming $3\,\,\mathrm{ab}^{-1}$ of integrated luminosity. The results for the baseline selections described in the preceding section are shown in Figs.~\ref{fig:master_MV_gp} and \ref{fig:master_MN_V2_LHC}; it is evident that DV searches in $V\rightarrow NN$ are not only poised to discover the RHN predicted by the see-saw mechanism, but that these searches may actually be the primary discovery mode for new gauge interactions with $M_V\lesssim400$ GeV, surpassing even the most optimistic projection for sensitivity to the dilepton resonance channel. The RHN parameter space accessible by such a search is far removed from the projected sensitivity of any other current experiment, as shown in Fig.~\ref{fig:master_MN_V2_LHC}.

\begin{figure}[t]
\centering
\includegraphics[width=0.5 \textwidth ]{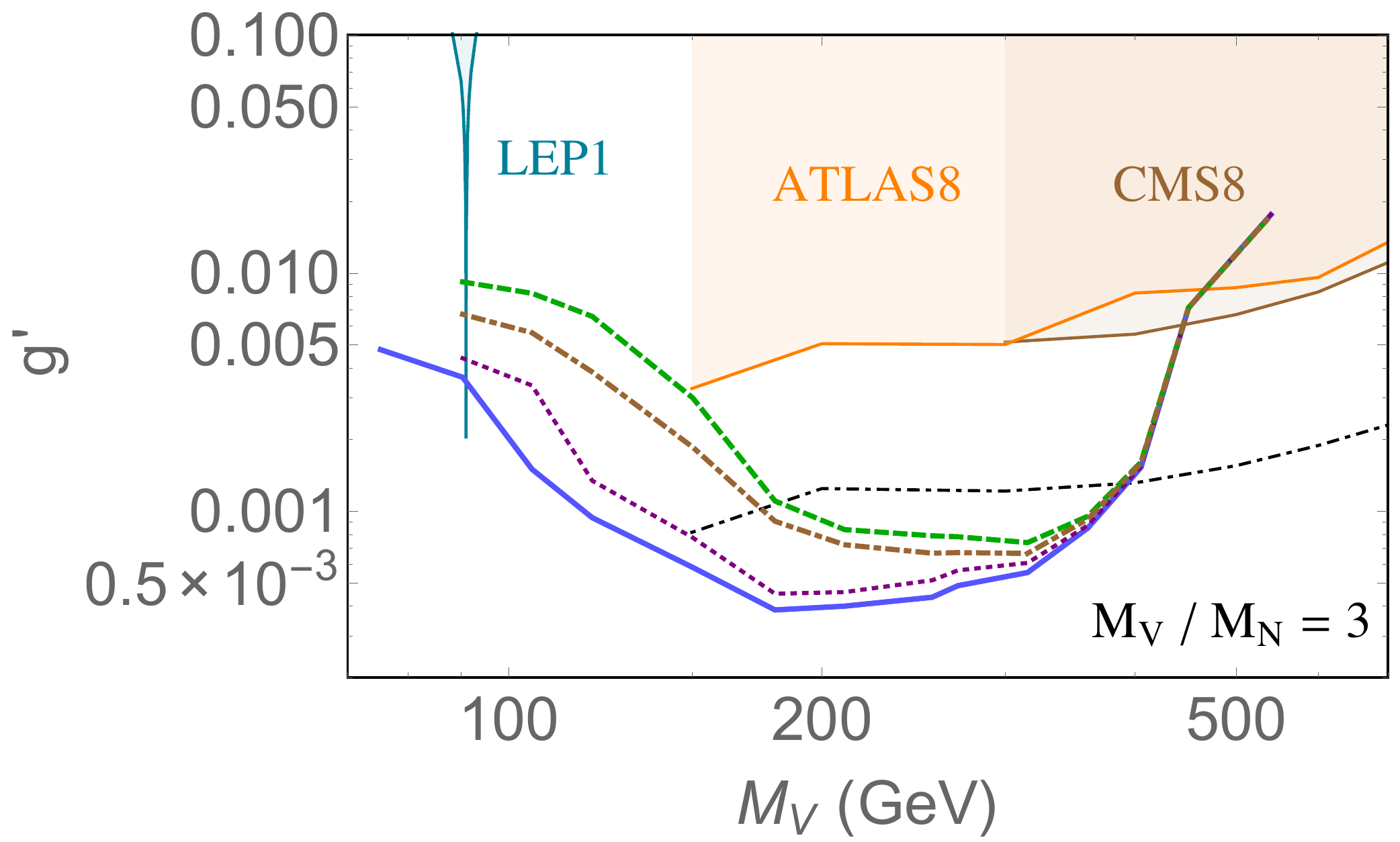}
\caption{Projected sensitivity to $V\rightarrow NN$ in searches for pairs of displaced objects at the high-luminosity LHC $(M_V/M_N=3)$. The sensitivity  is shown for different signal efficiency working points:~the baseline selection (blue solid), higher threshold trigger (purple dotted), more pessimistic vertex tagging efficiency (brown thick dot-dashed), and higher threshold trigger with pessimistic vertex tagging efficiency (green dashed); signal selections are described in the text. For comparison, the projected reach of the HL-LHC to $V\rightarrow\ell^+\ell^-$ is also shown (black dot-dashed). The RH neutrino mixing angle is fixed using Eq.~(\ref{thetass}). }
\label{fig:proj_2lepton_DV_MV_gp}
\end{figure}

\begin{figure}[t]
\centering
\includegraphics[width=0.5 \textwidth ]{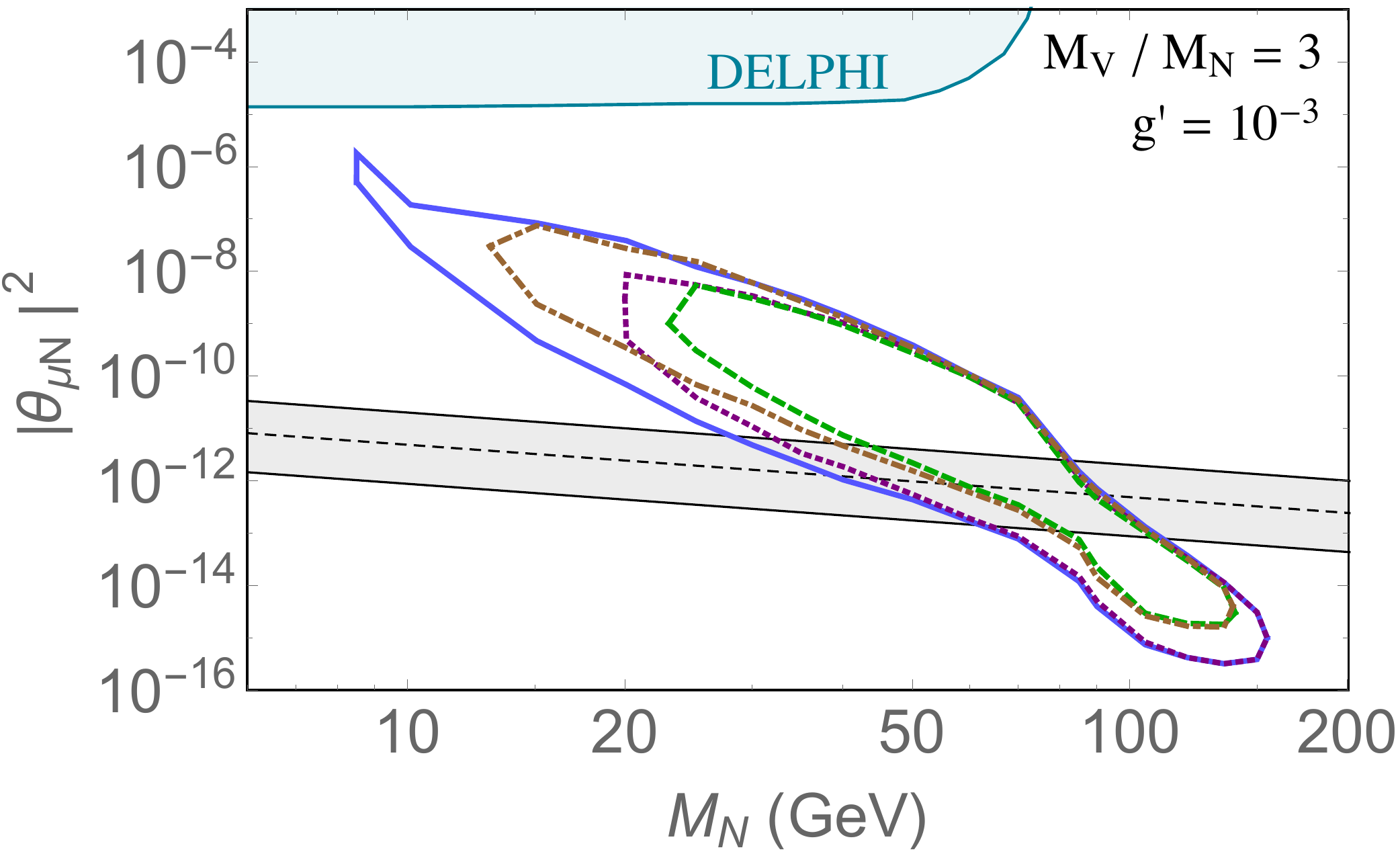}
\caption{Projected sensitivity to RHN parameters from searches for pairs of displaced objects at the high-luminosity LHC ($M_V/M_N=3$, $g'=10^{-3}$). The sensitivity  is shown for different signal efficiency working points as described in Fig.~\ref{fig:proj_2lepton_DV_MV_gp}. }
\label{fig:pproj_2lepton_DV_MN_V2}
\end{figure}

To assess the dependence of our results on the trigger and vertex-reconstruction assumptions made in our baseline selection, we also show the projected sensitivity for searches with  higher trigger thresholds and/or more pessimistic vertex reconstruction efficiencies described above. These results are shown in Figs.~\ref{fig:proj_2lepton_DV_MV_gp}-\ref{fig:pproj_2lepton_DV_MN_V2}; the results are qualitatively similar to the baseline selection and continue to have sensitivity to unexplored parameter space. Higher trigger thresholds worsen sensitivity to small $M_V$ 
since only events with hard initial state radiation pass the higher threshold  trigger, while higher masses are unaffected. Because the more pessimistic tagging efficiencies penalize object reconstruction at larger decay length, the HL-LHC sensitivity is worse at long lifetime (or, equivalently, small $|V_{\mu N}|^2$ and low $M_N$) with these selections. This is clearly seen in Fig.~\ref{fig:pproj_2lepton_DV_MN_V2}.

\begin{figure}[t]
\centering
\includegraphics[width=0.5 \textwidth ]{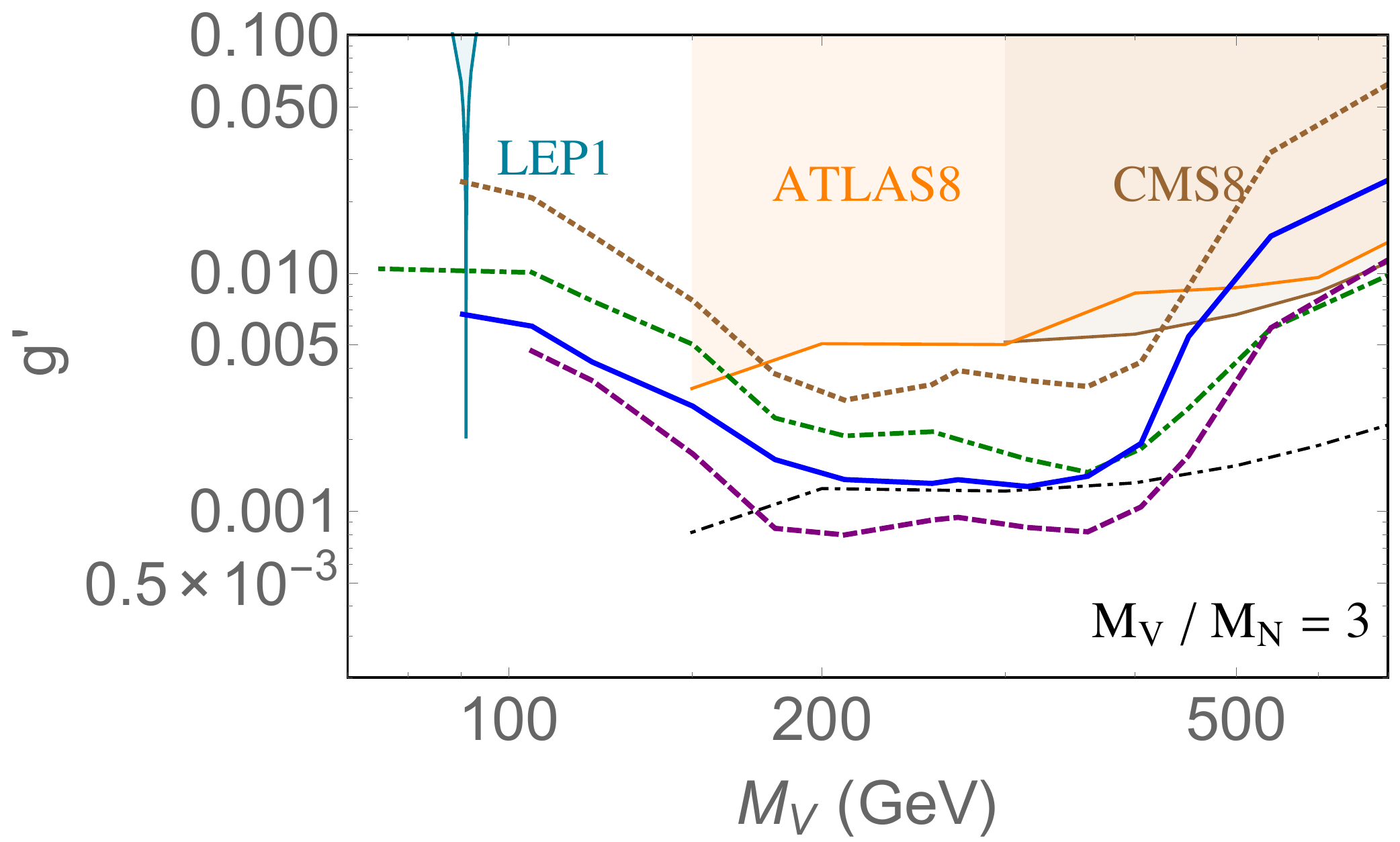}
\caption{Extrapolation to the high-luminosity LHC of current displaced vertex search strategies. Sensitivity is shown to $V\rightarrow NN$ $(M_V/M_N=3)$. Searches considered are the ATLAS displaced dilepton vertex search (blue, solid) \cite{Aad:2015rba}; ATLAS displaced muon + tracks vertex search (brown, dotted) \cite{Aad:2015rba}; CMS displaced dilepton vertex search (green, dot-dashed) \cite{CMS:2014hka}; a variant of the CMS displaced dilepton search without vertex requirement (purple, dashed) \cite{Khachatryan:2014mea}.  For comparison, the projected reach of the HL-LHC to $V\rightarrow\ell^+\ell^-$ is also shown (black dot-dashed). The RH neutrino mixing angle is fixed using Eq.~(\ref{thetass}). }
\label{fig:extr_2lepton_DV_MV_gp}
\end{figure}

We also compare the results for our analysis to extrapolations of the current Run 1 searches. To make a fair comparison, we assume that upgrades to the detector are sufficient to keep backgrounds low and show curves for sensitivity to five signal events. All efficiencies are kept the same as the existing analyses. We do make two changes to one analysis:~in the CMS ``displaced supersymmetry'' analysis \cite{Khachatryan:2014mea}, we additionally include same-flavor lepton pairs\footnote{This is motivated, in part, by the observation that the backgrounds for displaced $e\mu$ vertices is comparable to that for $\mu\mu$ \cite{Aad:2015rba}. Without reconstructing a common vertex, cosmic rays become more of a concern for events with two muons, but as the cosmic rate is independent of instantaneous luminosity, this background should remain manageable; we impose the same cosmic veto as in Ref.~\cite{CMS:2014hka}. Ref.~\cite{Evans:2016zau} found similarly small backgrounds for displaced $\mu^+\mu^-$.} and extend the vertex acceptance in $|d_0|$ out to 20 cm, consistent with other CMS analyses \cite{CMS:2014hka}.  We show the results in Fig.~\ref{fig:extr_2lepton_DV_MV_gp}; the CMS ``displaced supersymmetry'' is the most powerful, but appears not to quite rival our proposed 2-DV analysis in part because of the veto of events with more than two leptons and the requirement that the leptons be of the opposite sign, which reduces signal efficiency to the RHN model. The other searches do have sensitivity to currently unexplored parameter space, but face competition from the HL-LHC dilepton resonance searches. These results show some of the limitations of current  searches and the prospects for analyses that are optimized to the $V\rightarrow NN$ signal by requiring two displaced objects while simultaneously relaxing other selections to improve signal efficiency.

Finally, we comment that our proposed analysis exploits only one of the many signals associated with pairs of RHN decay. Other signatures that we have not studied in detail include fully hadronic DVs and missing energy signatures in conjunction with displaced leptons. While the typical momentum of these objects may be relatively low, the sensitivity to the $B-L$ model may be improved relative to our results shown in Figs.~\ref{fig:proj_2lepton_DV_MV_gp}-\ref{fig:pproj_2lepton_DV_MN_V2} by combining the results from multiple channels. In the event of the discovery of a signal, the relative population of leptonic and hadronic decay modes could provide valuable evidence to distinguish the RHN model presented here from other new physics scenarios. It may also be possible to exploit lepton-number-violating signals to discern the Majorana nature of the RHN (see also Refs.~\cite{Dicus:1991fk,Datta:1993nm,Almeida:2000pz,Panella:2001wq,Han:2006ip,delAguila:2007qnc,Perez:2008zc,Dev:2013wba,Alva:2014gxa,Das:2015toa}).
\\

\noindent {\bf Muon Spectrometer Searches:}~Due to the challenges of simulating vertex reconstruction in the MS, we only extrapolate  Run 1 results to the HL-LHC; we require two hadronic DVs in the MS and apply trigger and vertex reconstruction efficiencies from the ATLAS analysis \cite{Aad:2015uaa}. We choose the efficiencies for the $m_{\pi_V}=25$ GeV scenario in Ref.~\cite{Aad:2015uaa} Hidden Valley model because, of the efficiencies shown, it has the lowest-mass long-lived state and  best represents the relatively low-mass $N$ decays in our model. Nevertheless, we truncate our results at $3M_N=M_V>20$ GeV to avoid extrapolating the ATLAS results into the low-mass regime where we have no comparison of efficiencies. 

Our projections for the MS analysis are shown in Figs.~\ref{fig:master_MV_gp} and \ref{fig:master_MN_V2_LHC}; in doing so, we consider two background scenarios. In one, we assume that the Run 1 observed background of two events scales linearly with luminosity (along with an additional factor of two to approximately account for the higher energy of collisions) and show the $2\sigma$ signal sensitivity assuming only statistical uncertainties; this corresponds to approximately 50 signal events at the HL-LHC. We also show sensitivity to five signal events under the optimistic assumption that improvements to detectors and/or tracking can suppress the backgrounds. The improved reach shows the motivation for developing new methods for suppressing backgrounds at the HL-LHC if possible.

%% file: SHiP.tex
Beam dump experiments can provide a complementary probe of light RHNs and new gauge bosons:~while their limited center-of-mass energy restricts their sensitivity to $M_V\lesssim10$ GeV, their high rate of collision allows them to probe much smaller couplings than are possible at the LHC. 
One example is the proposed SHiP experiment at CERN~\cite{Alekhin:2015byh}, which would direct the energetic Super Proton Synchrotron (SPS) proton beam onto a target
of high density material, and use  muon shielding to extinguish any fluxes of SM particles other than neutrinos. 
In the target, the light RHN and/or $B-L$ gauge bosons can be abundantly produced in the proton-nucleus collisions and, if long-lived, can travel a macroscopic distance and eventually decay downstream in a detector to visible SM final states. Such a setup is highly efficient at probing light RHNs with masses in the (sub-)GeV range and decay lengths of order the target-detector separation distance.

RHNs that are directly produced through their mixing with SM neutrinos (\emph{i.e.,}~those with no new gauge interactions)  are prime targets and motivations for the SHiP experiment. The sensitivity of SHiP to such RHNs, which are produced and decay via the weak interaction, has been computed in Ref.~\cite{Alekhin:2015byh}, and we show this sensitivity in Fig.~\ref{fig:master_MN_V2_SHiP}. SHiP will be able to explore a significant range of new parameter space, including RHN masses up to the $B$-meson threshold of $M_N \lesssim {\cal O}(5 \,  {\rm GeV})$ and mixing angles down to $|\theta|^2 \gtrsim {\cal O}(10^{-9})$. While this reach is indeed impressive, it still appears challenging to probe the well-motivated parameter region obeying the see-saw relation in Eq.~(\ref{thetass}).

Here, we  estimate the sensitivity of the SHiP experiment to RHNs in the gauged $B-L$ scenario; since the production rate depends only on the new gauge coupling, $g'$, SHiP can be sensitive to much smaller mixing angles than would  otherwise be possible. We consider QCD production of $B-L$ vector bosons, 
$pp\rightarrow V$, followed by the prompt decay $V\rightarrow NN$. This results in a flux of $N$ particles emerging from the target, assumed here to be composed of Molybdenum. A fraction of these $N$ particles will pass through the detector and decay to visible final states, which can be detected by SHiP. 
The total event rate is given by 
\begin{equation}
N_{\rm evt} = 2  \, X_{NN} \,  N_{\rm POT} \,  {\rm Br}_{N,{\rm vis}} \, \epsilon_{\rm dec}.  
\label{eq:SHiP-estimate}
\end{equation}  
Here, $X_{NN} \equiv \sigma_{NN}/\sigma_{p{\rm Mo}}$, is the production fraction of $NN$ pairs (\emph{i.e.,}~the number of $NN$ pairs produced per proton on target), with $\sigma_{NN} \equiv \sigma(pp\rightarrow V \rightarrow NN)$ being the $NN$ production cross section, and $\sigma_{p{\rm Mo}} \simeq 10.7$ mb is the total proton-Molybdenum target cross section per target nucleon. Furthermore, $N_{\rm POT} = 4.5 \times 10^{20}$ is the number of protons on target (POT) proposed to be delivered to the SHiP experiment, $ {\rm Br}_{N,{\rm vis}}$ is the branching ratio of $N$ to visible final states, and $\epsilon_{\rm dec}$ is the probability of a produced $N$ particle to decay in the detector region. For the purposes of our calculation, we consider ``visible final states'' to be any decay mode of $N$ that produces some visible particles in the detector; restricting our analysis to fully reconstructible decay modes would give a somewhat reduced, but qualitatively similar, sensitivity to the one we compute. 

The various factors entering into Eq.~(\ref{eq:SHiP-estimate}) are computed as follows. The $NN$ production cross section is given by
\begin{equation}
\sigma_{NN} = \frac{\pi g'^2 }{27 M_V^2}  \sum_{q} {\cal F}_{q\bar q}(\tau) \, {\rm Br}_{V \rightarrow NN},
\end{equation}
where ${\cal F}_{q\bar q}(\tau)$ is the parton luminosity,
\begin{equation}
{\cal F}_{q\bar q}(\tau) \equiv \tau \int_{\tau}^1 \frac{dx}{x} \left[\,f_q(x) f_{\bar q}(\tau/x) +f_{\bar q}(x) f_{q}(\tau/x)  \,\right],
\end{equation}
with $f_i(x)$ the parton distribution function for parton $i$ (we employ the NNPDF2.3LO PDF set \cite{Ball:2014uwa}),  $\tau \equiv M_V^2/s$, and $\sqrt{s} \simeq \sqrt{2 m_p E_{\rm SPS}} \approx 27$ GeV with $E_{\rm SPS} = 400$ GeV for the CERN SPS proton beam. Furthermore, ${\rm Br}_{V \rightarrow NN}$ is the branching ratio of the $B-L$ gauge boson to $NN$, and is approximately $10\%$ as discussed in Section \ref{sec:simplified_model}.
For instance, fixing $g' = 10^{-4}$, and $M_V = 3 M_N$, we find a cross section $\sigma_{NN} \approx 10$ fb ($3\times 10^{-3}$  fb) for $m_V = 2$ GeV (10 GeV). 
The production fraction $X_{NN}$ in Eq.~(\ref{eq:SHiP-estimate}) follows straightforwardly from $\sigma_{NN}$ as discussed above. Furthermore, the branching ratio of $N$ to visible final states, ${\rm Br}_{N,{\rm vis}}$, is computed according to the weak decay partial widths provided in Ref.~\cite{Gorbunov:2007ak}. 
Finally, to compute the acceptance factor, $\epsilon_{\rm dec}$, we have performed a Monte Carlo simulation and generated $NN$ events using \texttt{MadGraph5\_aMC@NLO} \cite{Alwall:2014hca}. For each simulated event $i$ in which the $N$ passes through the detector, we compute the probability $\epsilon_i$ for it to decay within the detector according to the formula
\begin{equation}
\epsilon_i  = {\rm exp}\left( - \frac{\ell_{1,i}}{\gamma_i \beta_i \tau_N }\right) - {\rm exp}\left( - \frac{\ell_{2,i}}{\gamma_i \beta_i \tau_N }\right), 
\end{equation}
where $\ell_{1}$ ($\ell_2$) is the distance from the target to the point of entry (exit), $\gamma_i$ ($\beta_i$) is the Lorentz boost factor (velocity) of the $N$ particle, and $\tau_N$ is the RHN lifetime. 
From the MC simulation, we obtain 
\begin{equation}
\epsilon_{\rm dec} = \frac{\sum_i \epsilon_i}{N_{\rm gen}},
\end{equation}
where $N_{\rm gen}$ is the total number of generated $N$ events. 

Given that SHiP is designed to be a nearly background free experiment,  we estimate a Poisson 95$\%$ C.L.~sensitivity, $N^{95}_{\rm evt} = 3$ events. 
In Figure~\ref{fig:master_MV_gp}, we show the sensitivity of SHiP in the $M_V- g'$ plane, fixing $M_N = M_V/3$, and $\theta$ according to the see-saw relation in Eq.~(\ref{thetass}). Currently, the strongest constraint in the $M_V = 1-10$ GeV range comes from BaBaR and BESIII searches for $e^+ e^- \rightarrow \gamma V \rightarrow \gamma \ell^+ \ell^-$, and extends down to couplings of order $g' \sim 3 \times 10^{-4}$ for BaBar (and below $10^{-4}$ for some masses from BESIII). We observe that SHiP will be sensitive to RHN production from $V\rightarrow NN$ for couplings that are smaller than the current BaBar limits by a factor of a few, corresponding to roughly an order of magnitude improvement in the $B-L$ fine structure constant $\alpha'$; its sensitivity would be comparable to the reach of Belle II in the dilepton channel. To show the  sensitivity of SHiP to RHN parameters, we  fix $g' = 10^{-4}$ and $M_V = 3 M_N$, displaying the results in the $M_N - |\theta|^2$ plane in Figure.~\ref{fig:master_MN_V2_SHiP}. In this case, we see that SHiP's sensitivity extends well beyond a number of existing constraints and can probe down to the see-saw motivated region for masses $M_{N}\sim\mathcal{O}(\mathrm{GeV})$. We also observe the enhanced sensitivity in this model compared to RHN's produced through the  decays of heavy-flavor mesons.

%% file: concl.tex
Right-handed neutrinos ($N$) are some of the best-motivated candidates for extensions of the SM as they can account for the observed SM neutrino masses via the see-saw mechanism. However, the smallness of the SM neutrino masses suggests that $N$ are very feebly coupled to SM fields if they are within  kinematic reach of current experiments, $M_N\lesssim$ TeV. This makes their direct study at colliders and beam-dump experiments  very difficult.

In this paper, we have explored the discovery prospects for $N$ in current and planned experiments where there exist enhanced interactions between $N$ and the SM. Instead of considering modifications of the neutrino mass matrices that would allow for larger mixing between $N$ and the SM neutrinos, we study the scenario where there exists an additional mediator that couples $N$ to the SM, giving pair production at accelerator and collider experiments. We have concentrated on the case of a new ``dark force'', namely a $B-L$ gauge interaction with coupling constants smaller than those of the SM gauge groups; because three RHNs are needed to cancel the chiral anomalies in the new gauge interaction, these models naturally incorporate new RHN interactions. We have shown that high-energy colliders (such as the LHC) and beam-dump experiments (such as SHiP) have excellent sensitivity to the pair production of $N$ through the $B-L$ gauge interaction, and the subsequent displaced decays of $N$; remarkably, current and upcoming experiments can have sensitivity to the tiny mixing angles between SM neutrinos and $N$ motivated by the see-saw mechanism. We have also demonstrated that long-lived RHN signatures can serve as a primary discovery mode for new feebly coupled gauge interactions, giving sensitivity to $B-L$ gauge couplings that are too small for detection in other experiments.

Because the see-saw mechanism suggests that RHNs decay on macroscopic distances only for $M_N\lesssim200$ GeV, much of the sensitivity of experiments to these models is in the low-mass regime, well below the hadronic centre-of-mass energy of the LHC. It is therefore crucial that momentum thresholds for LHC searches remain low in high-luminosity running to retain sensitivity to RHNs, which may necessitate modifications to existing search strategies such as requiring an additional displaced object to suppress backgrounds. While  we have focused only on a few displaced decay modes of $N$ in our LHC study, the LHC could obtain even better sensitivity by combining all possible RHN decay modes; in the event of a signal, this would allow the experiments to distinguish the Dirac or Majorana nature of RHNs as well as to disentangle the flavor structure of the RHN sector.

RHNs are the stated main physics target for the SHiP facility. We have shown that in models with additional gauge interactions of 
RHN, the sensitivity of SHiP is complemented by the projected reach of the high-intensity electron-positron colliders. This way, the 
GeV scale dark sector (RHNs and ``dark force'') could be discovered and studied at multiple facilities.

{\bf \emph{Acknowledgments}}: We would like to thank Valentin Hirschi, Eder Izaguirre, Zhen Liu, Stefan Prestel, and Brock Tweedie for useful discussions. 
This work was made possible by the facilities of the Shared Hierarchical Academic Research Computing Network (SHARCNET) and Compute/Calcul Canada. This research was supported in part by Perimeter Institute for Theoretical Physics. Research at Perimeter Institute is supported by the Government of Canada through Industry Canada and by the Province of Ontario through the Ministry of Research and Innovation.
This work was performed in part at the Aspen Center for Physics, which is supported by National Science Foundation grant PHY-1066293.